\documentclass[11pt,aps,prd,preprint,groupedaddress,tightenlines,nofootinbib,floatfix]{revtex4}
\usepackage[dvipdfmx]{graphicx}

\usepackage{amssymb}
\usepackage{amsmath}
\usepackage{epstopdf}
\usepackage{booktabs}

\makeatletter
\newcommand{\figcaption}[1]{\def\@captype{figure}\caption{#1}}
\newcommand{\tblcaption}[1]{\def\@captype{table}\caption{#1}}
\makeatother

\def\simge{\mathrel{%
       \rlap{\raise 0.511ex \hbox{$>$}}{\lower 0.511ex \hbox{$\sim$}}}}
\def\simle{\mathrel{
       \rlap{\raise 0.511ex \hbox{$<$}}{\lower 0.511ex \hbox{$\sim$}}}}

\begin{document}

\title{Multipoint reweighting method and its applications to lattice QCD}

\author{R. Iwami$^{1}$, S.~Ejiri$^{2}$, K.~Kanaya$^{3,4}$, Y.~Nakagawa$^{1}$, 
D.~Yamamoto$^{1}$, and T.~Umeda$^{5}$ \\
(WHOT-QCD Collaboration)
}
\affiliation{
$^1$Graduate School of Science and Technology, Niigata University, Niigata 950-2181, Japan\\
$^2$Department of Physics, Niigata University, Niigata 950-2181, Japan\\
$^3$Center for Integrated Research in Fundamental Science and Technology (CiRfSE), University of Tsukuba, Tsukuba, Ibaraki 305-8571, Japan\\
$^4$Faculty of Pure and Applied Sciences, University of Tsukuba, Tsukuba, Ibaraki 305-8571, Japan\\
$^5$Graduate School of Education, Hiroshima University, Hiroshima 739-8524, Japan
}
\date{October 14, 2015}

\begin{abstract}
The reweighting method is widely used in numerical studies of QCD,
in particular, for the cases in which the conventional Monte-Carlo method
cannot be applied directly, e.g., finite density QCD.
However, the application range of the reweighing method is restricted 
due to several problems.  
One of the most severe problems here is the overlap problem.
To solve it, we examine a multipoint reweighting method in which
simulations at several simulation points are combined in the data analyses.
We systematically study the applicability and limitation of the multipoint 
reweighting method in two-flavor QCD at zero density.
Measuring histograms of physical quantities at a series of simulation
points, we apply the multipoint reweighting method to calculate
the meson masses as continuous functions of the gauge coupling
$\beta$ and the hopping parameters $\kappa$.
We then determine lines of constant physics and beta functions,
which are needed in a calculation of the equation of state at
finite temperature.
\end{abstract}

\maketitle

\section{Introduction}
\label{introduction}

At extremely high temperatures and/or densities, the quark matter is expected to turn into new phases. 
Clarification of the nature of these states as well as the phase structure of QCD is important in understanding the evolution of the Universe around microseconds to milliseconds after the big bang. 
Here, the only method to obtain information about the quark matter directly from the first principles of QCD is to numerically study QCD based on Monte Carlo simulations on the lattice. 
In the study of lattice QCD with dynamical quarks, the Boltzmann weight is proportional to the quark determinant. 
At nonzero chemical potentials, however, the quark determinant becomes complex and causes a serious problem -- the Monte Carlo procedure is not justified because of the complex Boltzmann weight.
In the low density region, because the fluctuations of the complex phase are small, we can avoid the problem by the reweighting method in which the complex phase is treated as a correction factor of the observables (reweighting factor). 
Using the reweighting method, we can also vary coupling parameters of the system by absorbing the difference of the Boltzmann weight at different coupling parameters into the reweighting factor. 
This is powerful in a study of the phase structure in which a survey over a range of coupling parameter space is mandatory.

At higher densities, however, larger fluctuations of the complex phase introduce two severe difficulties.
One is the sign problem. 
Because of large fluctuations of the complex phase in the reweighting factor, exponentially large statistics is required to obtain reliable estimates for the reweighted observables.
Several methods have been proposed to remedy or mitigate the sign problem \cite{signproblem,whot10}.
Another problem is the overlap problem.
When we try to shift simulation parameters largely by the reweighting method, the reweighting factor tries to enhance part of the Boltzmann weight whose statistical quality is low.
This can be easily seen by viewing the Boltzmann factor in terms of the histogram for relevant observables.
When expectation values of observables vary largely with the shift of the simulation parameters, 
the reweighting factor has to enhance part of the histogram far from the original peak position.
Because it is statistically quite hard to achieve highly accurate details of the histogram around such a point, the reweighting method may lead to completely unreliable results (see, e.g. ref.~\cite{whot14}).
This makes it difficult to study high density QCD, in which the transition is expected to be of first order and thus expectation values can jump largely around the transition point. 

In this paper, we focus on the overlap problem. 
The overlap problem is expected to be milder if one changes a couple of parameters at the same time. 
In an early trial to identify the critical point in a high density region of QCD
\cite{FK,BS02,Ejiri04}, Fodor and Katz shifted the chemical potential and the gauge coupling (temperature) simultaneously along the crossover curve to achieve a better overlap (multiparameter reweighting method).
More recently, the WHOT-QCD Collaboration investigated the phase structure of $N_{\rm f}$-flavor QCD in the heavy-quark region 
and found that the system at large quark masses is controlled by only two combinations of parameters, $\beta +48 \sum_{f=1}^{N_{\rm f}} \kappa_f^4$ and $\sum_{f=1}^{N_{\rm f}} \kappa_f^{N_t} \cosh (\mu_f/T)$, 
where $\beta=6/g^2$ is the gauge coupling, $\kappa_f$ and $\mu_f$ are the hopping parameter and chemical potential for the $f$ th flavor, and $N_t$ is the temporal lattice size \cite{whot14}.
This means that, when one changes the coupling parameters while keeping these combinations constant, the system does not change and thus the overlap problem does not arise.
We expect that similar combinations of parameters also exist in the light-quark region.

In the study of ref.~\cite{whot14}, the multipoint reweighting method \cite{FS89} for $\beta$ played an important role: 
Combining configurations obtained at different $\beta$, we could calculate the effective potential in a wide range of the observable values, which was mandatory in a reliable evaluation of the transition point.

In the present paper, we extend the multipoint reweighting method to the multiparameter space of $\beta$ and $\kappa_f$, and test if the method helps to overcome the overlap problem in the light-quark region,  
performing simulations in a simpler case of zero-density QCD.
We measure histograms of physical quantities at a series of simulation points in two-flavor QCD 
and apply the multipoint reweighting method to calculate the meson masses as continuous
functions of $\beta$ and $\kappa$.
We then determine lines of constant physics in 
the $(\beta, \kappa)$ space and evaluate the derivatives of the
lattice spacing with respect to $\beta$ and $\kappa$ along
the lines of constant physics (inverse of the beta functions), which are needed in a calculation
of the equation of state at finite temperature.

In the next section, the multipoint reweighting method is introduced, and in Sec.\ref{sec:overlap}, we examine the overlap problem by performing numerical simulations in two-flavor QCD. 
We then calculate the meson masses, the lines of constant physics, and the derivatives of the lattice spacing with respect to $\beta$ and $\kappa$ along the lines of constant physics in Sec.~\ref{sec:beta-func}. 
Section \ref{sec:conclusion} contains our conclusions.

\section{Multipoint reweighting method}
\label{sec:method}

\subsection{Multiparameter reweighting method}

Let us consider QCD with $N_{\rm f}$ flavors of quarks and define a histogram for a set of physical quantities 
$X = (X_1,X_2,\cdots)$ by 
\begin{eqnarray}
w(X; \beta, \kappa, \mu) 
&=& \int {\cal D} U \ \prod_i \delta(X_i - \hat{X}_i) \ 
e^{-\hat{S}_G}\ \prod_{f=1}^{N_{\rm f}} \det \hat{M}(\beta, \kappa_f, \mu_f) .
\label{eq:dist}
\end{eqnarray}
where $\hat{S}_G$ is the gauge action, $\hat{M}$ is the kernel matrix of the quark action, and 
$\hat{X} = (\hat{X}_1,\hat{X}_2,\cdots)$ are the operators for $X$.
The coupling parameters of the theory are the gauge coupling $\beta$, the hopping parameters  
$(\kappa_1, \kappa_2, \cdots, \kappa_{N_{\rm f}})$, and the chemical potentials 
$(\mu_1, \mu_2, \cdots, \mu_{N_{\rm f}})$. 
For simplicity, we denote the set of coupling parameters $(\beta, \kappa_1,\cdots, \mu_1,\cdots)$ as $b$.

Then, the partition function is given by
$ Z(b)  = \int\! w(X;b) \, dX $
with $dX = \prod_i dX_i$, 
and the probability distribution function of $X$ is given by $Z^{-1}(b)\, w(X;b)$. 
The expectation value of an operator ${\cal O} [\hat{X}]$, which is written in terms of the operators $\hat{X}$ is evaluated as 
\begin{eqnarray}
\langle {\cal O }[\hat{X}] \rangle_{(b)} = \frac{1}{Z(b) } 
\int\! {\cal O} [X] \, w(X; b) \, dX .
\label{eq:expop}
\end{eqnarray}
For convenience, we also define the effective potential as
\label{eq:veff}
\begin{equation}
V_{\rm eff}(X; b) = -\ln w(X; b) .
\end{equation}

Let us consider a calculation of the histogram at $b$ using configurations generated at $b_0$.
Such a calculation can be easily done with the reweighting method by choosing $S(b)$ and $S(b_0)$ as the first two elements of $X$,
where
\begin{eqnarray}
\hat{S} = \hat{S}_G - \sum_f \ln \det \hat{M}
\end{eqnarray}
is the effective action of QCD 
and $S(b)$ is the value of the action with the coupling parameters $b$ evaluated on the configuration generated at the simulation point.
Let us denote $S(b) \equiv S$ and $S(b_0) \equiv S_0$, and redefine $X$ as the set of remaining elements of $X$ other than $S$ and $S_0$. 
The histogram obtained by the simulation at $b_0$ is given by $w(X,S,S_0;b_0)$. 
From (\ref{eq:dist}), we find that the histogram at $b$ is simply given by 
\begin{equation}
w(X, S,S_0; b) 
= e^{-(S-S_0)} \, w(X, S, S_0; b_0),
\label{eq:chbeta}
\end{equation}
and the histogram of $X$ at $b$ is given by 
\begin{eqnarray}
w(X; b) = 
\int w(X, S, S_0; b) \, dS \,dS_0 .
\label{eq:wint}
\end{eqnarray}

In a simple case of quenched QCD, $\hat{S}=-6 N_{\rm site} \beta \hat{P}$ with $\hat{P}$ the plaquette and $N_{\rm site}$ the number of sites (lattice volume). 
Then, because both $S$ and $S_0$ are fixed when $P$ is fixed, we have $ w(P; \beta_0)= w(P, S, S_0; \beta_0)$, and 
\begin{eqnarray*}
w(P; \beta) &=& e^{6 N_{\rm site}(\beta - \beta_0)P} \, w(P; \beta_0) , 
\\
V_{\rm eff}(P; \beta) &=& V_{\rm eff}(P; \beta_0) - 6 N_{\rm site}(\beta - \beta_0)P .
\end{eqnarray*}
If we approximate $w(P; \beta_0)$ by a Gaussian distribution centered at $\bar{P}_0 = \langle \hat{P} \rangle_{(b_0)}$, i.e., 
$w(P;\beta_0) = \exp[-V_{\rm eff} (P;\beta_0)] \propto \exp[-\alpha (P - \bar{P}_0)^2]$ 
with an appropriate constant $\alpha$, 
then the expectation value of $\hat{P}$ at $\beta$ is given by 
$\bar{P} = \bar{P}_0 + 3 N_{\rm site}(\beta - \beta_0)/\alpha$. 
When $\beta - \beta_0$ is large, $\bar{P}$ leaves 
the statistically reliable region of the original histogram $w(P;\beta_0)$ and thus the results at $\beta$ become unreliable (the overlap problem).
As mentioned in the Introduction, in order to study the expected first-order transition of QCD at high densities, we need to obtain $w$ and $V_{\rm eff}$ reliably in a wide range of $X$.

\subsection{Multipoint multiparameter reweighting method}

To overcome the overlap problem, we extend the reweighting formulas to combine configurations obtained 
at different simulation points \cite{whot14,FS89} for the case with dynamical quarks. 
We perform a series of $N_{\rm sp}$ simulations at $b_i$ with the number of 
configurations $N_i$ where $i=1, \cdots , N_{\rm sp}$.
Let us denote $\vec{S} = (S_1,\cdots,S_{N_{\rm sp}})$ with  $S_i=S(b_i)$.
Using (\ref{eq:chbeta}), the probability distribution function at $b_i$ is related to the histogram at $b$ as  
\begin{eqnarray}
Z^{-1}(b_i) \,w(X,S, \vec{S}; b_i) = Z^{-1}(b_i) \, e^{-(S_i -S)} \,w(X,S, \vec{S}; b) 
\end{eqnarray}
where $S=S(b)$.
We then obtain 
\begin{eqnarray}
\sum_{i=1}^{N_{\rm sp}} N_i \, Z^{-1}(b_i) \, w(X, S, \vec{S}; b_i) 
= e^{S} 
\sum_{i=1}^{N_{\rm sp}} N_i \, Z^{-1}(b_i) \, e^{-S_i}  \, w(X, S, \vec{S}; b).
\label{eq:sum1}
\end{eqnarray}
Note that the left-hand side of (\ref{eq:sum1}) gives the naive histogram using all the configurations disregarding the difference in the simulation parameters $b_i$. 
From this relation, we find
\begin{eqnarray}
w(X,S, \vec{S}; b)= G(S, \vec{S};b,\vec{b}) \,
\sum_{i=1}^{N_{\rm sp}} N_i \, Z^{-1}(b_i) \, w(X,S, \vec{S}; b_i) 
\end{eqnarray}
where $\vec{b}=(b_1,\cdots,b_{N_{\rm sp}})$ and 
\begin{eqnarray}
G(S, \vec{S};b,\vec{b})=\frac{ e^{-S}}{
\sum_{i=1}^{N_{\rm sp}} N_i \, e^{-S_i} Z^{-1}(b_i)} .
\end{eqnarray}
This expression means that the histogram $w(X,S, \vec{S};b)$ at $b$ is given by multiplying $G(S, \vec{S};b,\vec{b})$ by the naive histogram.

To calculate $G(S, \vec{S};b,\vec{b})$, we need the values of $Z(b_i)$, the partition function at $b_i$.
Here, we note that the partition function $Z$ at $\beta$ is given by
\begin{eqnarray}
Z(b) &=& \int w(X,S, \vec{S}; b) \,dXdSd\vec{S} 
 \nonumber\\
&=&
 \sum_{i=1}^{N_{\rm sp}} N_i \int G(S,\vec{S};b,\vec{b}) \, Z^{-1}(b_i) \, w(X,S, \vec{S}; b_i) \,dXdSd\vec{S}
 \nonumber\\
&=&
\sum_{i=1}^{N_{\rm sp}} N_i \left\langle G(\hat{S},\vec{\hat{S}};b,\vec{b}) \right\rangle_{\!(b_i)}, 
\end{eqnarray}
which is just the naive sum of $G(S,\vec{S};b,\vec{b})$ over all the configurations disregarding the difference in the simulation parameters.
Then, $Z(b_i)$ can be determined by the consistency relations, 
\begin{eqnarray}
Z(b_i) 
=\sum_{k=1}^{N_{\rm sp}} N_k \left\langle G(\hat{S},\vec{\hat{S}};b_i,\vec{b}) \right\rangle_{\! (b_k)}
=\sum_{k=1}^{N_{\rm sp}} N_k \left\langle 
\frac{e^{-\hat{S}_i}}{
\sum_{j=1}^{N_{\rm sp}} N_j e^{-\hat{S}_j} Z^{-1}(b_j)} \right\rangle_{\! (b_k)}
\end{eqnarray}
for $i=1,\cdots,N_{\rm sp}$, but up to an overall factor. 
Denoting $f_i=-\ln Z(b_i)$, these equations can be rewritten as
\begin{eqnarray}
1 = \sum_{k=1}^{N_{\rm sp}} N_k \left\langle
\left( \sum_{j=1}^{N_{\rm sp}} N_j \exp[\hat{S}_i -\hat{S}_j - f_i +f_j] \right)^{-1} 
\right\rangle_{\! (b_k)}.
\hspace{5mm}
i=1,\cdots,N_{\rm sp}.
\label{eq:consis}
\end{eqnarray}
Starting from appropriate initial values of $f_i$, we solve (\ref{eq:consis}) numerically by an iterative method. 
Note that one of the $f_i$'s must be fixed to remove the ambiguity corresponding to the undetermined overall factor.

The expectation value of an operator $\hat{X}$ at $b$ can be evaluated as
\begin{eqnarray}
\langle \hat{X} \rangle_{(b)} 
\;=\; \frac{1}{Z(b)} \int X \, w(X,S, \vec{S}; b) \, dS \,d\vec{S}
\;=\; \frac{1}{Z(b)} \sum_{i=1}^{N_{\rm sp}} N_i \left\langle \hat{X} \, G(\hat{S},\vec{\hat{S}};b,\vec{b}) \right\rangle_{\!(b_i)} ,
\label{eq:multibeta}
\end{eqnarray}
which is just the naive sum of $X G$ over all the configurations disregarding the difference in the simulation parameters.
From this formula, we see that the histogram of $X$ at $b$ is given by
\begin{eqnarray}
w(X;b) 
\;=\; \sum_{i=1}^{N_{\rm sp}} N_i \left\langle \delta(X- \hat{X}) 
\, G(\hat{S},\vec{\hat{S}};b,\vec{b}) \right\rangle_{\!(b_i)}. 
\label{eq:multihis}
\end{eqnarray}

\section{Test study of overlap problem}
\label{sec:overlap}

\subsection{Two-flavor QCD}
\label{sec:model}

To test the multipoint reweighting method, we perform simulations of QCD with degenerate two-flavor clover-improved Wilson quarks and RG-improved Iwasaki glues at zero density.
The gauge action is given by 
\begin{eqnarray}
\hat{S}_G = -6 N_{\rm site} \,\beta \, \hat{P},
\label{eq:Sg}
\end{eqnarray} 
where 
$N_{\rm site} = N_s^3 \times N_t$ is the lattice volume,
and $\hat{P}$ is the improved plaquette by Iwasaki,
\begin{eqnarray}
\hat{P} = c_0 \hat{W}^{(1 \times 1)} + 2c_1 \hat{W}^{(1 \times 2)},
\end{eqnarray} 
with $c_1=-0.331$, $c_0=1-8c_1$ \cite{Iwasaki}, and $\hat{W}^{i \times j}$ is the $(i \times j)$ Wilson loop.
The quark action is given by
\begin{eqnarray}
\hat{S}_Q &=& \sum_{f=1}^2 \sum_{x,y} \bar{\psi}^f_x \, \hat{M}_{xy} \, \psi^f_y ,
\label{eq:Sq}
\\
\hat{M}_{xy} &=& \delta_{xy}-\kappa
\sum_\mu \left\{ (1-\gamma_\mu)\,\hat{U}_{x,\mu}\delta_{x+\hat{\mu},y} + 
(1+\gamma_\mu)\,\hat{U}^\dagger_{x-\hat{\mu},\mu}\delta_{x-\hat{\mu},y}
\right\}
- \delta_{xy}\, c_{\rm SW}\, \kappa \sum_{\mu>\nu}\sigma_{\mu\nu}
\hat{F}_{x\mu\nu}
\nonumber
\end{eqnarray} 
where 
$\kappa$ is the hopping parameter common to two flavors, 
and $\hat{F}_{x\mu\nu}$ is the standard clover-shaped lattice field strength. 
For the clover coefficient $c_{SW}$, we adopt a mean-field value by   
substituting the one-loop result for the plaquette, $c_{SW}=(1-0.8412 \beta^{-1})^{-3/4}$.
The improvement parameters of the action are the same as those adopted in refs.~\cite{whot07,whot10,CPPACS00,CPPACS01}.

The simulations are carried out on an $8^4$ lattice at 9 simulation points 
(all the combinations of $\beta= \{ 1.800$, 1.825, 1.850\} and $\kappa= \{ 0.1400$, 0.1425, 0.1440\}) 
for the test study in this section, and 
on a $16^4$ lattice at 30 points 
(all the combinations of $\beta= \{ 1.806$, 1.8125, 1.819, 1.825, 1.831, 1.837\} and 
$\kappa= \{ 0.14000$, 0.14125, 0.14250, 0.14300, 0.14400\})
for the determination of lines of constant physics and beta functions in Sec.~\ref{sec:beta-func}.
The number of configurations for the measurement is 200 at each simulation point, and statistical errors are estimated by a jackknife method.

\subsection{Reweighting with dynamical quarks}
\label{sec:lndetM}

\begin{figure}[t]
\centering
\includegraphics[width=80mm]{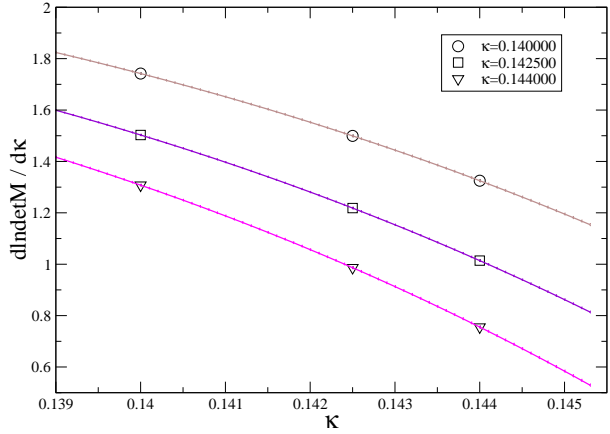}
\caption{The average of 
$\partial \ln \det M/ \partial \kappa$ and its cubic spline interpolations on each configuration generated at $\beta=1.80$ and $\kappa=0.1400$ (circle), $0.1425$ (square) and $0.1440$ (triangle).
Statistical errors are estimated to be around the thickness of the curves.
}
\label{fig:spline}
\end{figure}

\begin{figure}[t]
\centering
\includegraphics[width=85mm]{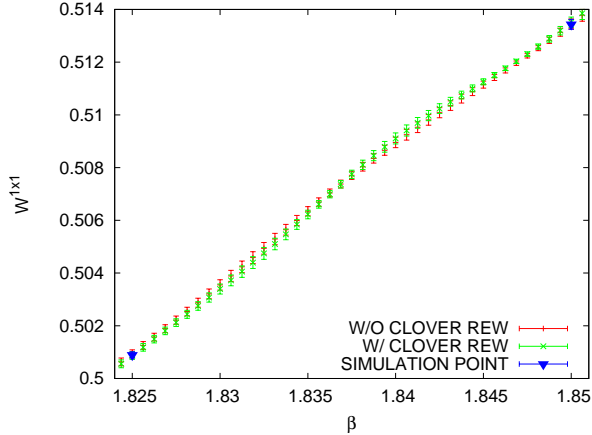}
\caption{Effect of the $\beta$-dependence in $c_{SW}$ on the value of $1\times1$ Wilson loop at $\kappa=0.140$. 
Two filled triangles represent the results obtained directly at the simulation points $\beta=1.825$ and 1.850.
The green and red curves, which are almost completely overlapping with each other, are the results of the reweighting with and without the linear term in (\ref{eq:clover}).}
\label{fig:clover}
\end{figure}

The effective action $S = S_G - \sum_f \ln \det M$ consists of the gauge part $S_G$ and the quark part $\ln \det M$.
Calculation of the latter requires large computational cost.
In this study, we evaluate it by measuring the first and second $\kappa$ derivatives of $\ln \det M$ 
at several $\kappa_i$'s, where $i=1, 2, \cdots$, on each configuration,  
and interpolate $\ln \det M$ between $\kappa_i$ and $\kappa_{i+1}$ 
assuming a quadratic function in terms of $\kappa$
\footnote{We note that $\ln \det M$ is an even function of $\kappa$. However, the first derivative exists at nonvanishing $\kappa$, and the expansion (\ref{eq:lndetMquad}) is possible assuming cancellation of odd-power terms in $\kappa$ by the higher order terms.
Although we can consider an alternative expansion in terms of, e.g., $(\kappa^2 - \kappa_i^2)$, the difference is absorbed by the higher order terms, and the quality of the fit is not improved in the present case.
Because the derivatives in $\kappa$ are directly related to measurable observables, we prefer the expansion (\ref{eq:lndetMquad}).
},
\begin{eqnarray}
\ln \det M(\kappa) = \ln \det M(\kappa_i) +C_1 (\kappa -\kappa_i) +C_2 (\kappa -\kappa_i)^2 
+C_3 (\kappa -\kappa_i)^3 +C_4 (\kappa -\kappa_i)^4.
\label{eq:lndetMquad}
\end{eqnarray}
Then, the derivatives are written as
\begin{eqnarray}
\frac{\partial \ln \det M}{\partial \kappa} (\kappa) 
&=& C_1 +2 C_2 (\kappa -\kappa_i) +3 C_3 (\kappa -\kappa_i)^2 +4 C_4 (\kappa -\kappa_i)^3, \\
\frac{\partial^2 \ln \det M}{\partial \kappa^2} (\kappa) 
&=& 2 C_2 +6 C_3 (\kappa -\kappa_i) +12 C_4 (\kappa -\kappa_i)^2. 
\end{eqnarray}
We fix the four coefficients $C_a$ such that the first and second derivatives 
reproduce the measured values at $\kappa_i$ and $\kappa_{i+1}$, i.e., 
\begin{eqnarray}
C_1 = d^{(1)}_i, \  
C_2 = \frac{1}{2} d^{(2)}_i, \ 
C_3 = \frac{d^{(1)}_{i+1} - d^{(1)}_i}{h^2} 
- \frac{d^{(2)}_{i+1} +2 d^{(2)}_i}{3h}, \ 
C_4 = -\frac{d^{(1)}_{i+1} - d^{(1)}_i}{2h^3} 
+ \frac{d^{(2)}_{i+1} + d^{(2)}_i}{4h^2}.
\end{eqnarray}
where $d^{(1)}_i = [\partial \ln \det M / \partial \kappa] (\kappa_i)$, 
$d^{(2)}_i = [\partial^2 \ln \det M / \partial \kappa^2] (\kappa_i)$ and 
$h = \kappa_{i+1} - \kappa_i$.
Using the coefficients $C_a$, we obtain $\ln \det M (\kappa)$ parameterized by (\ref{eq:lndetMquad})
in the range of $\kappa_i \leq \kappa \leq \kappa_{i+1}$.
This determines $\ln \det M$ up to an overall constant that is redundant in the reweighting calculations.

The derivatives of $\ln \det M$ are given by traces of some combination of $M^{-1}$ and derivatives of $M$ 
(see ref.~\cite{whot10} for explicit expressions).
We compute these traces by the random noise method.
To reduce errors due to finite number of noise vectors, $N_{\rm noise}$, 
we use the random noise method only for the trace over spatial indices, and we calculate the traces over color and spinor indices exactly.
As discussed in ref.~\cite{whot10}, this procedure helps us reduce $N_{\rm noise}$, in particular, with Wilson-type quarks.
In this study, we adopt $N_{\rm noise}=5$.

In Fig.~\ref{fig:spline}, we show an example of the interpolation for $\partial \ln \det M/ \partial \kappa$. 
Open symbols are the averages of this derivative measured at $\beta=1.80$ and $\kappa=0.1400$ (circle), $0.1425$ (square) and $0.1440$ (triangle). 
The curves (with error bars) are the results of interpolation using the data of $\partial \ln \det M/ \partial \kappa$ and $\partial^2 \ln \det M/ \partial \kappa^2$ with $\kappa=0.1400$, 0.1425 and 0.1440 on each configuration. 

The reweighting in the $\beta$ direction requires care because the quark kernel $M$ is dependent on $\beta$ through the clover coefficient $c_{SW}$ in our choice. 
We take into account the effect of the $\beta$ dependence in $c_{SW}$ by a linear approximation \cite{CswBeta}
\begin{eqnarray}
\ln \det M(\beta, \kappa)= \ln \det M(\beta_0, \kappa) 
+ (\beta - \beta_0) \, \left[ \frac{d c_{SW}}{d \beta} \frac{\partial \ln \det M}{\partial c_{SW}} \right]_{\beta_0, \kappa} .
\label{eq:clover}
\end{eqnarray}
In Fig.~\ref{fig:clover}, we show the results of $1\times1$ Wilson loop at $\kappa=0.140$ with and without the linear term in (\ref{eq:clover}). 
We find that the differences are at most 0.05\% and are much smaller than the statistical errors. 
We thus consider that the effects of the $\beta$ dependence in $c_{SW}$ is small and thus the linear approximation is sufficiently safe in the range of the coupling parameters we study.
In the following, we adopt the linear approximation (\ref{eq:clover}).

\subsection{Overlap problem and multipoint reweighting}

\begin{figure}[t]
\centering
\includegraphics[width=75mm]{improve_8888_MKREW_PoS_P_III.eps}
\hspace{5mm}
\includegraphics[width=63mm]{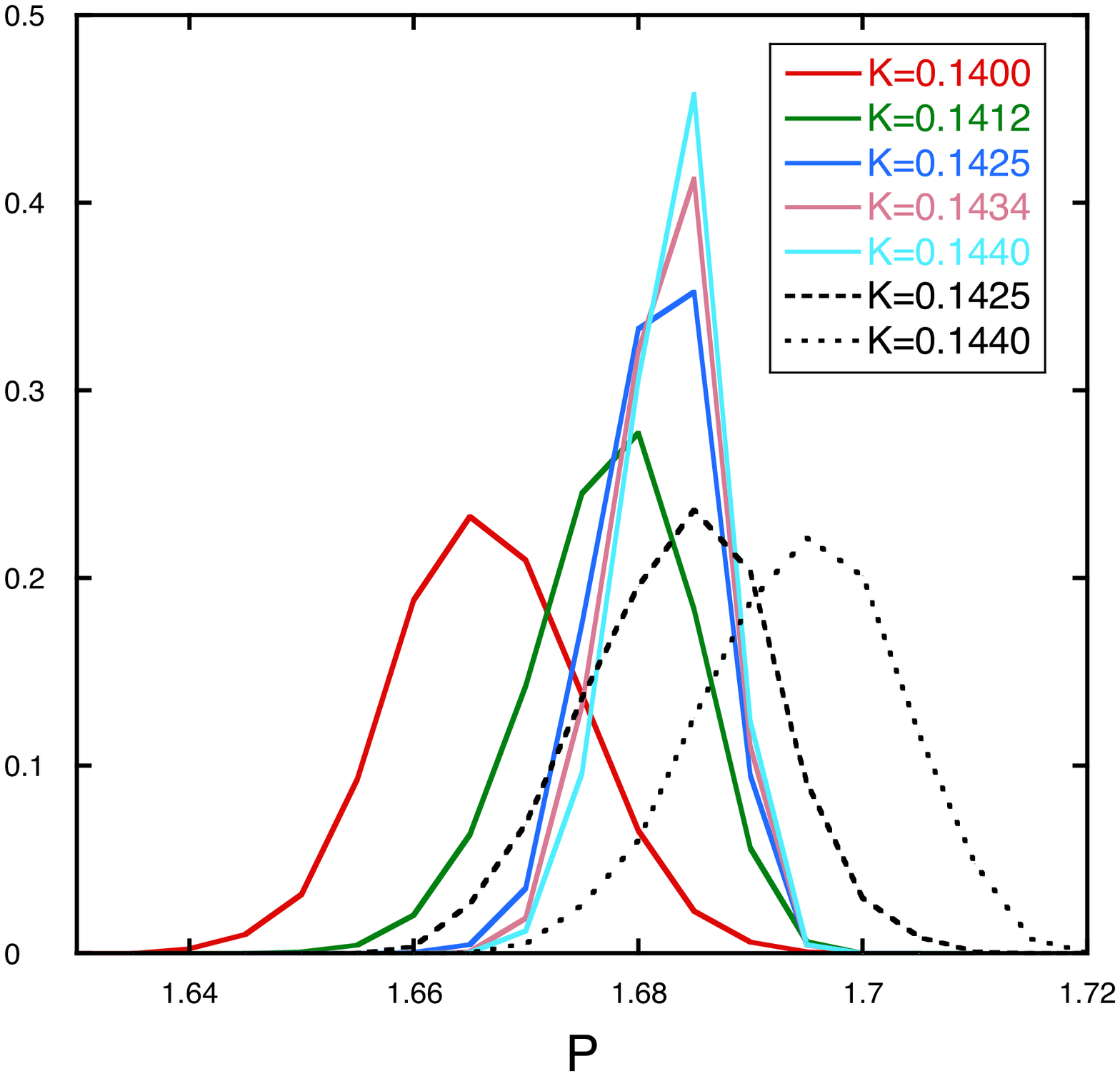}
\caption{Left: The expectation value of the improved plaquette $P \equiv c_0 W^{1 \times 1} + 2 c_1 W^{1 \times 2}$ at $\beta=1.825$. 
Black dots are the expectation values obtained by the simulations at $\kappa=0.1400$, 0.1425, and 0.1440.
Blue curve is the result of single-point reweighting using the configurations at $\kappa=0.140$ only.
Right: Red, green, blue, magenta and light blue curves are the probability distribution functions of $P$ at various $\kappa$'s obtained by the single-point reweighting method using the configurations at $\kappa=0.140$ only. 
Black dashed curves are the original probability distribution functions at the simulation points $\kappa=0.1425$ and 0.1440.}
\label{fig:overlap1}
\end{figure}

\begin{figure}[t]
\centering
\includegraphics[width=75mm]{improve_8888_MKREW_PoS_P.eps}
\hspace{5mm}
\includegraphics[width=63mm]{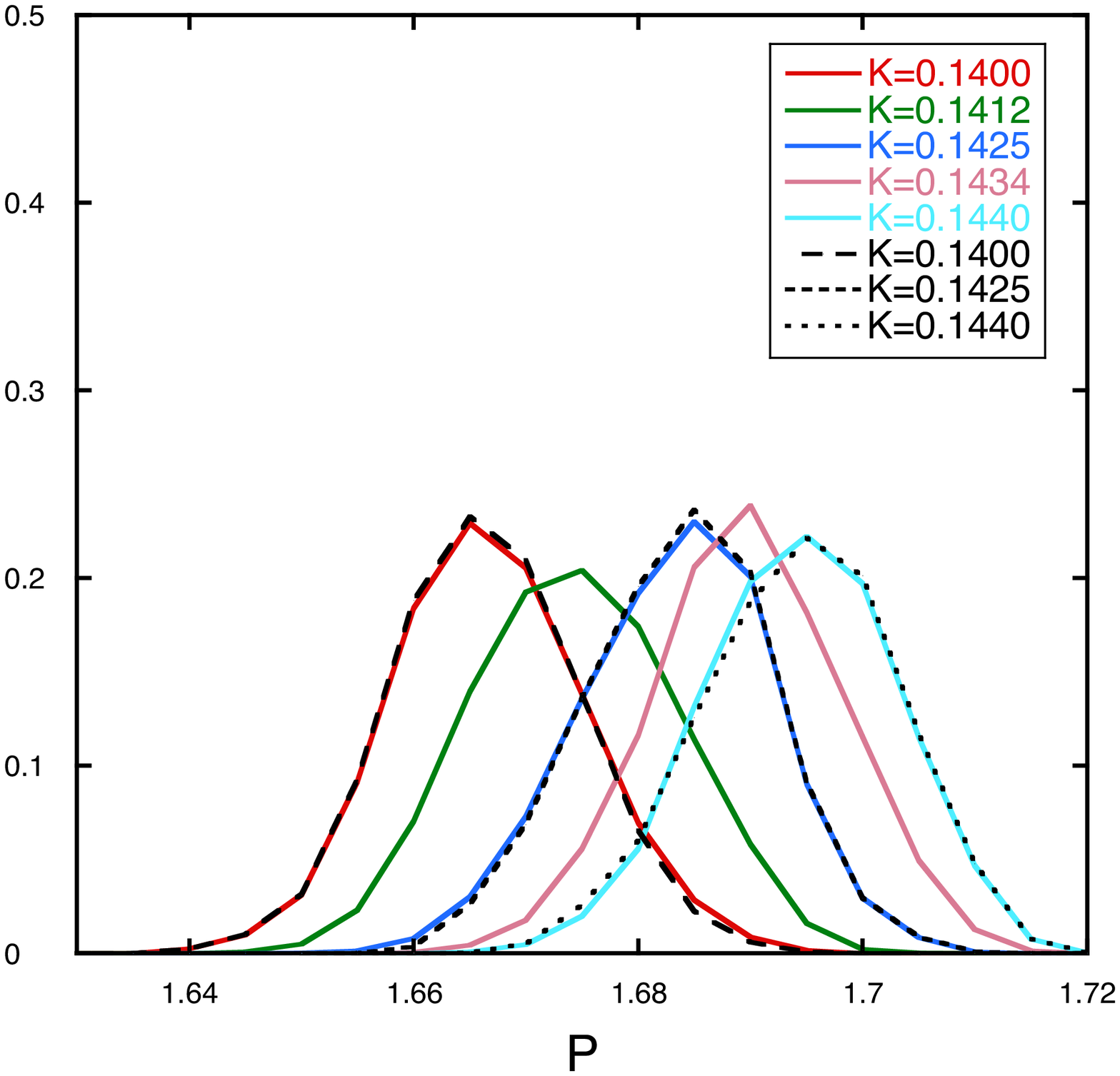}
\caption{Left: The expectation value of $P \equiv c_0 W^{1 \times 1} + 2 c_1 W^{1 \times 2}$ as a function of $\kappa$ at $\beta=1.825$.
Black dots are the expectation values obtained by the simulations at $\kappa=0.1400$, 0.1425, and 0.1440.
Blue, green and purple curves are the results of the single-point reweighting method 
using the configurations at $\kappa=0.1400$, 0.1425, and 0.1440, respectively. 
Red curve is the result of the multipoint reweighting method combining the configurations at the three $\kappa$'s.
Right: Red, green, blue, magenta and light blue curves are the probability distribution functions of $P$ by the multipoint reweighting method at various $\kappa$'s. 
Black dashed curves are the original probability distribution functions at the three simulation points, $\kappa=0.1400$, 0.1425, and 0.1440.}
\label{fig:overlap2}
\end{figure}

\begin{figure}[t]
\centering
\includegraphics[width=80mm]{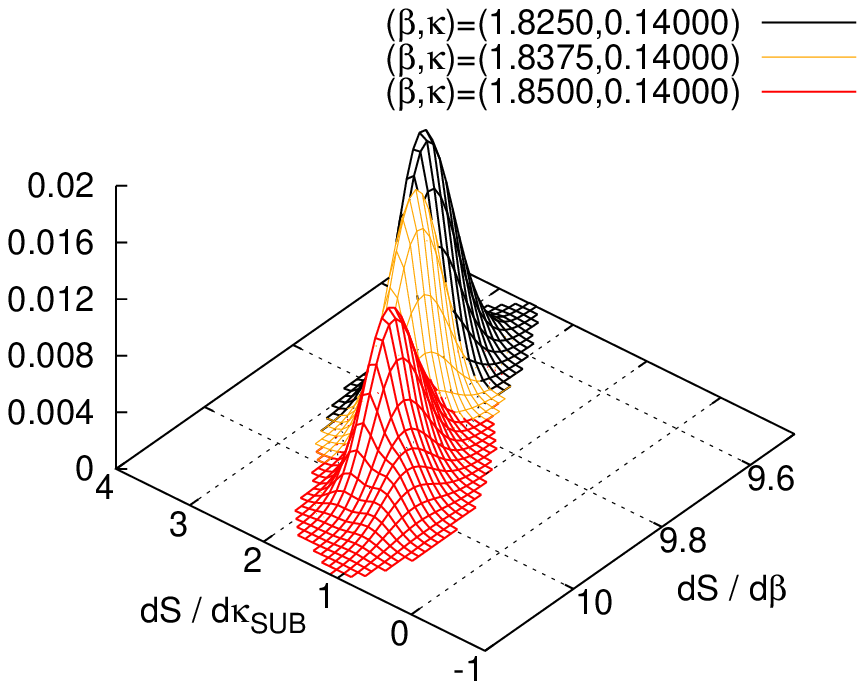}
\hspace{-5mm}
\includegraphics[width=80mm]{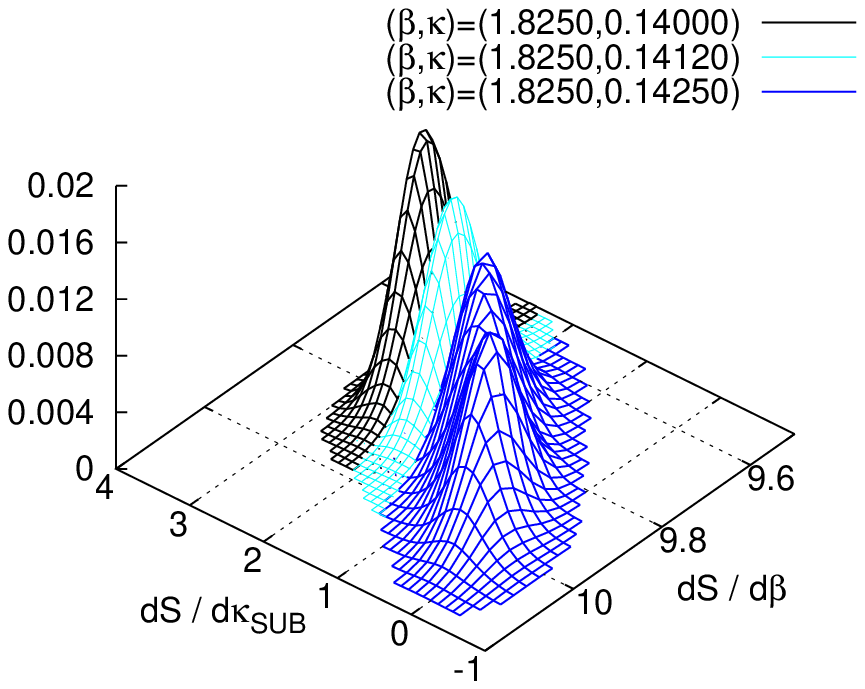}
\caption{The $\beta$-dependence (left) and $\kappa$-dependence (right) of the histogram 
for $N_{\rm site}^{-1} (\partial S/ \partial \beta)$ and 
$N_{\rm site}^{-1} [\partial S/ \partial \kappa]_{\rm SUB} \equiv 
N_{\rm site}^{-1} [\partial S/ \partial \kappa 
-(288 N_{\rm f} \kappa^4/c_0) (\partial S/ \partial \beta)]$, where $N_{\rm site}=8^4.$}
\label{fig:3Dplot}
\end{figure}

\begin{figure}[t]
\centering
\includegraphics[width=110mm]{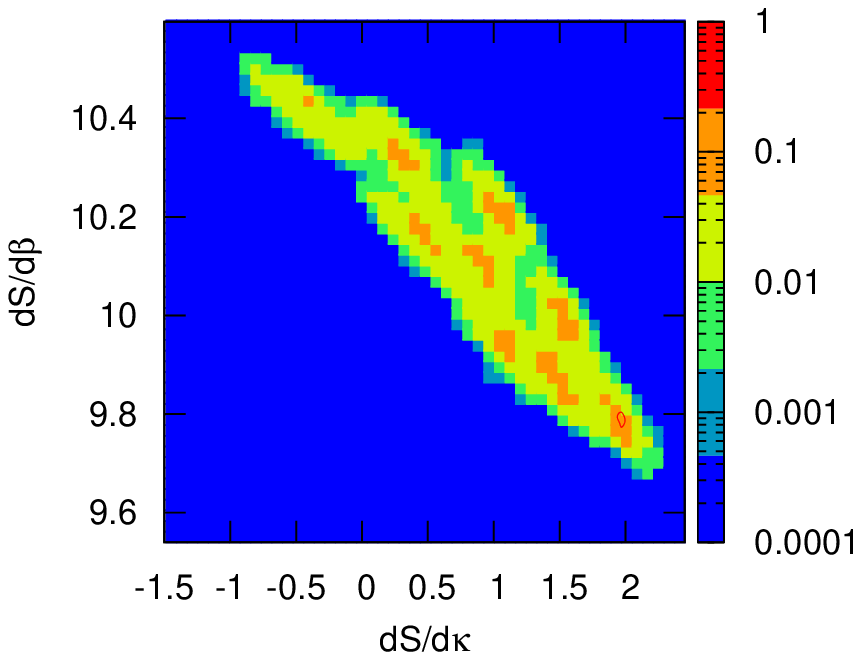}
\vspace{-5mm}
\caption{The histogram
of $N_{\rm site}^{-1} (\partial S/ \partial \beta)$ and 
$N_{\rm site}^{-1} (\partial S/ \partial \kappa)$ , 
normalized by the maximum height.
}
\label{fig:distr}
\end{figure}

In the left panel of Fig.~\ref{fig:overlap1}, we show the results for the improved plaquette
$P=c_0 W^{1 \times 1} + 2c_1 W^{1 \times 2}$ of the Iwasaki action at $\beta=1.825$. 
The black dots represent the expectation values of $P$ at the three simulation points without reweighting.
The blue curve shows the results of the single-point reweighting method using the configurations at $\kappa=0.1400$ only. 
We note that the blue curve fails to reproduce the data at $\kappa=0.1425$ and 0.1440. 
Even the error bars based on a standard jackknife analysis are unreliable. 
The reason can be easily understood by consulting the histogram of $P$:
The red curve in the right panel of Fig.~\ref{fig:overlap1} is the original probability distribution function at 
$(\beta, \kappa)=(1.825, 0.140)$, and green, blue, magenta and light blue curves are the probability distribution functions at $\kappa=0.1412$, 0.1425, 0.1434, and 0.1440, respectively, predicted by the single-point reweighting  (\ref{eq:chbeta}) using the configurations at $\kappa=0.1400$.
Because a probability distribution function at $\kappa$ other than the simulation point is calculated as a product of the reweighting factor and the original distribution function, 
the shifted distribution functions cannot go out of the range of the original distribution $P \sim 1.64$--1.695 at $\kappa=0.1400$, 
and they fail to reproduce the true distribution functions at $\kappa=0.1425$ and 0.1440 shown by the black dot-dashed curves.
Accordingly, the expectation value, which is approximately the peak position of the distribution function, cannot go out of the range of the original distribution, as shown in Fig.~\ref{fig:overlap2} (left).
Furthermore, due to the poor statistics around the boundary of the original distribution, we cannot estimate the errors there reliably.

The multipoint reweighting method introduced in the previous section can enlarge the range of reliable reweighting by combining different ranges of distribution obtained at different simulation points.
The result of multipoint reweighting combining the configurations at $\kappa=0.1400$, 0.1425, and 0.1440 is shown by a red curve in the left panel of Fig.~\ref{fig:overlap2}. 
Results of the single-point reweighting method using the configurations at $\kappa=0.1400$, 0.1425, and 0.1440 separately are also shown by blue, green and purple curves, respectively. 
We find that, unlike the case of single-point reweighting, the red curve smoothly connects all the simulation results with small errors. 
In the right panel of Fig.~\ref{fig:overlap2}, probability distribution functions from the multipoint reweighting method are plotted for $\kappa=0.1412$, 0.1424, 0.1436, and 0.144. 
The probability distribution functions reproduce and smoothly interpolate the original distribution functions at different simulation points.

The method is applicable to other observables too.
In the calculation of the equation of state, we calculate the following combination of energy density$(\epsilon)$ and pressure $(p)$,
\begin{eqnarray}
\frac{\epsilon - 3p}{T^4} 
&=& N_t^4 \left\langle \frac{1}{N_s^3 N_t} 
a \frac{d S}{d a} \right\rangle_{\! 0} \nonumber \\
&=& N_t^4 \left[
a \frac{d \beta}{d a} 
 \left\langle \frac{1}{N_s^3 N_t} 
 \frac{\partial S}{\partial \beta} \right\rangle_{\! 0}
+ a \frac{d \kappa}{d a} 
 \left\langle  \frac{1}{N_s^3 N_t} 
 \frac{\partial S}{\partial \kappa} \right\rangle_{\! 0} \, \right],
\label{eq:e3p}
\end{eqnarray}
where $\langle \cdots \rangle_0$ is the expectation value 
at finite temperature with the zero-temperature value subtracted. 
We then need the expectation values of the derivatives of the action 
$S = S_G - \sum_f \ln \det M$, i.e. 
$\partial S/ \partial \beta$ and $\partial S/ \partial \kappa$.

From our experience in the heavy quark region \cite{whot11}, we expect that the $\kappa$ dependences of $\partial S/ \partial \beta$ and $\partial S/ \partial \kappa$ are strongly correlated with each other. 
We thus consider the combination 
\begin{eqnarray}
\left[\frac{\partial S}{\partial \kappa}\right]_{\rm SUB} \equiv \frac{\partial S}{\partial \kappa} -\frac{288 N_{\rm f} \kappa^4}{c_0} \, \frac{\partial S}{\partial \beta}, 
\end{eqnarray}
as a component approximately perpendicular to  $\partial S/ \partial \beta$, 
by subtracting the leading order contribution in the hopping parameter expansion from $\partial S/ \partial \kappa$. 
In the right and left panels of Fig.~\ref{fig:3Dplot}, we show the results of the multipoint reweighting method for the $\beta$ and $\kappa$ dependence of the two-dimensional histogram of 
$\partial S/ \partial \beta$ and $[\partial S/ \partial \kappa]_{\rm SUB}$. 
The histograms are obtained by combining the configurations at 9 simulation points (3 $\beta$'s $\times$ 3 $\kappa$'s) on the $8^4$ lattice. 
We see that the histogram moves smoothly as $\beta$ and $\kappa$ are varied, without hitting a boundary.
We can thus compute the expectation values of $\partial S/ \partial \beta$ and $[\partial S/ \partial \kappa]_{\rm SUB}$ as continuous functions of $\beta$ and $\kappa$ in this range of the coupling parameters, without the overlap problem.

The applicable range of the multipoint reweighting method can be estimated easily from the sum of the histograms measured at each simulation point.
As explained in Sec.~\ref{sec:method}, the expectation values 
(\ref{eq:multibeta}) in the multipoint reweighting method are obtained by 
just the naive sum of the operators multiplied by the reweighting factor $G$ over 
all the configurations disregarding the difference of $\beta$ and $\kappa$.
We show, in Fig.~\ref{fig:distr}, the contour plot of the naive sum of the histograms obtained at all 9 simulation points in the $(\partial S/ \partial \beta, \partial S/ \partial \kappa)$ plane. 
In the region painted by bright colors, many configurations are available, and thus a reliable calculation is possible.

\section{Lines of constant physics and beta functions}
\label{sec:beta-func}

\begin{figure}[tb]
\centering
\includegraphics[width=80mm]{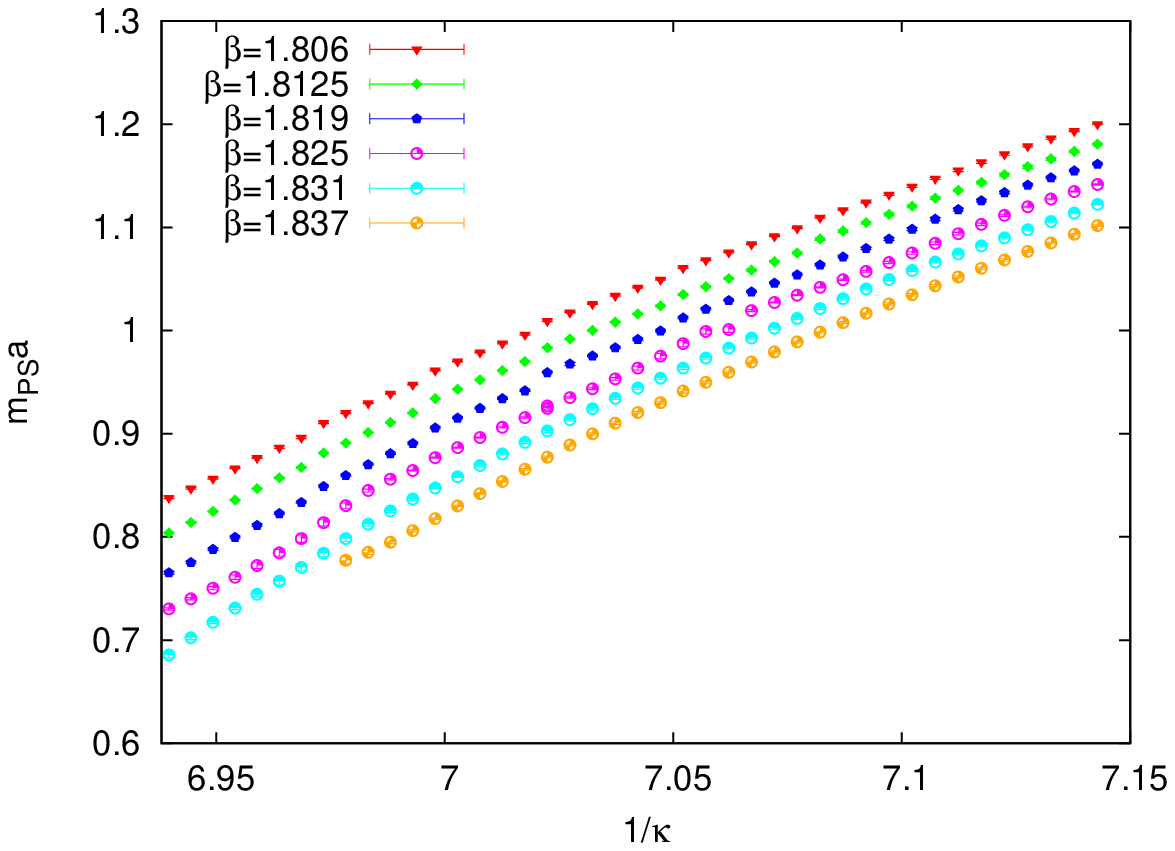}
\hspace{-2mm}
\includegraphics[width=80mm]{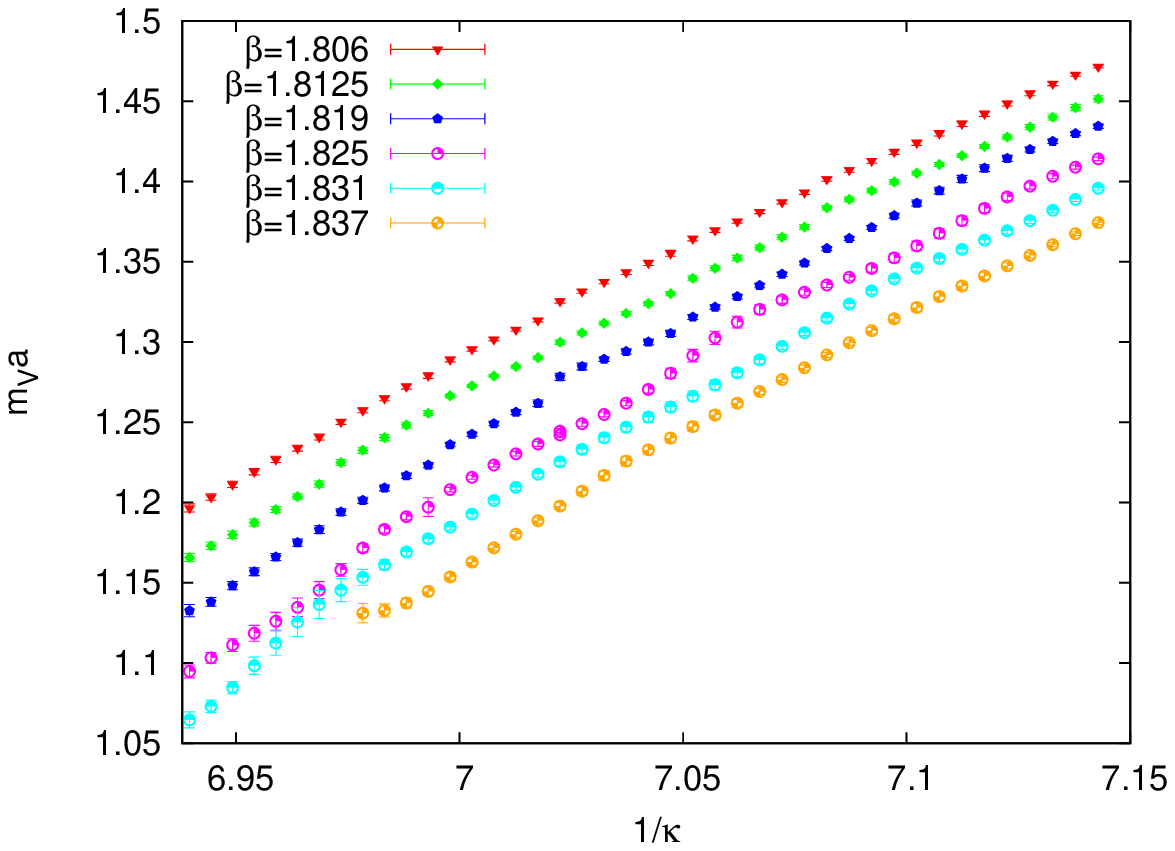}
\caption{The pseudoscalar meson mass $m_{\rm PS} a$ (left) and vector meson mass
$m_{\rm V} a$ (right) as functions of $1/\kappa$.  
Data points without a clear plateau are removed.
}
\label{fig:mass}
\end{figure}

\begin{figure}[tb]
\begin{minipage}{0.48\textwidth}
\includegraphics[width=80mm]{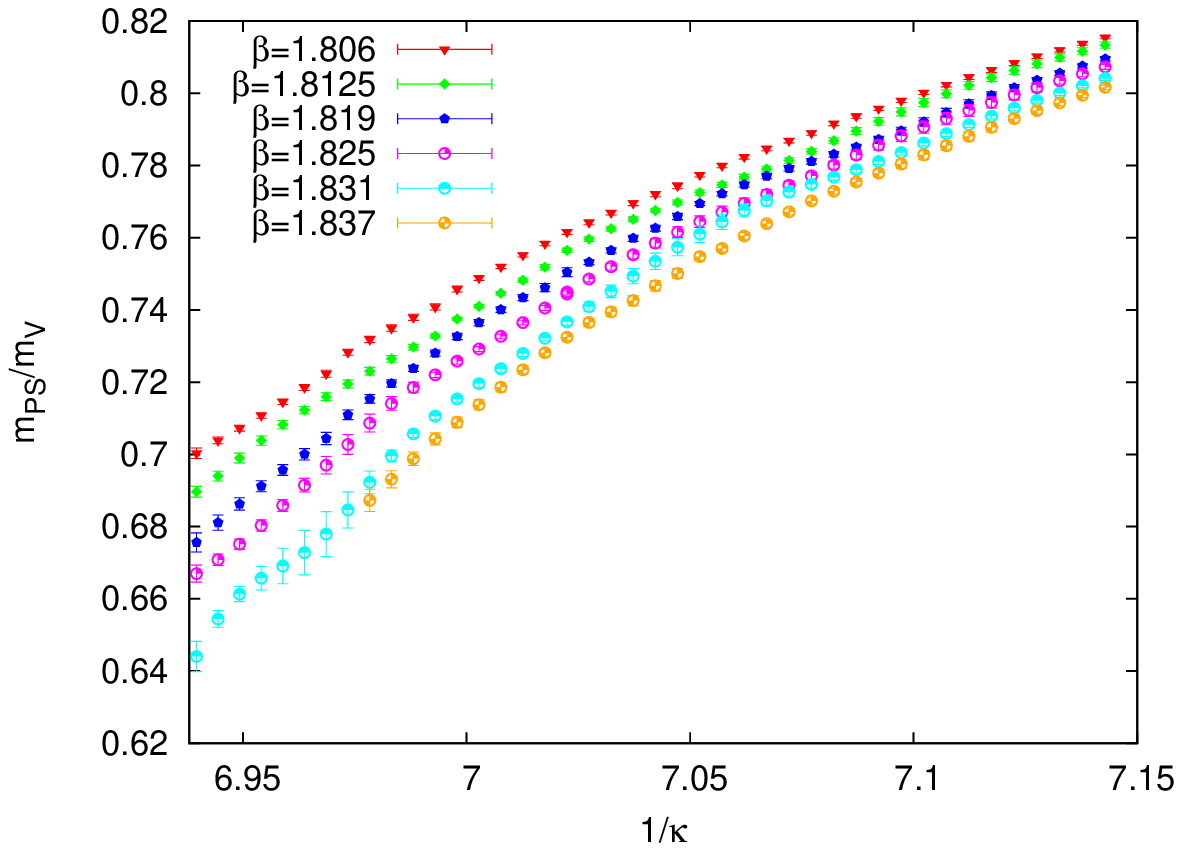}
\caption{The pseudoscalar and vector meson mass ratio $m_{\rm PS}/m_{\rm V}$ as functions of $1/\kappa$.  
}
\label{fig:pirho}
\end{minipage}
\hfill
\begin{minipage}{0.48\textwidth}
\includegraphics[width=80mm]{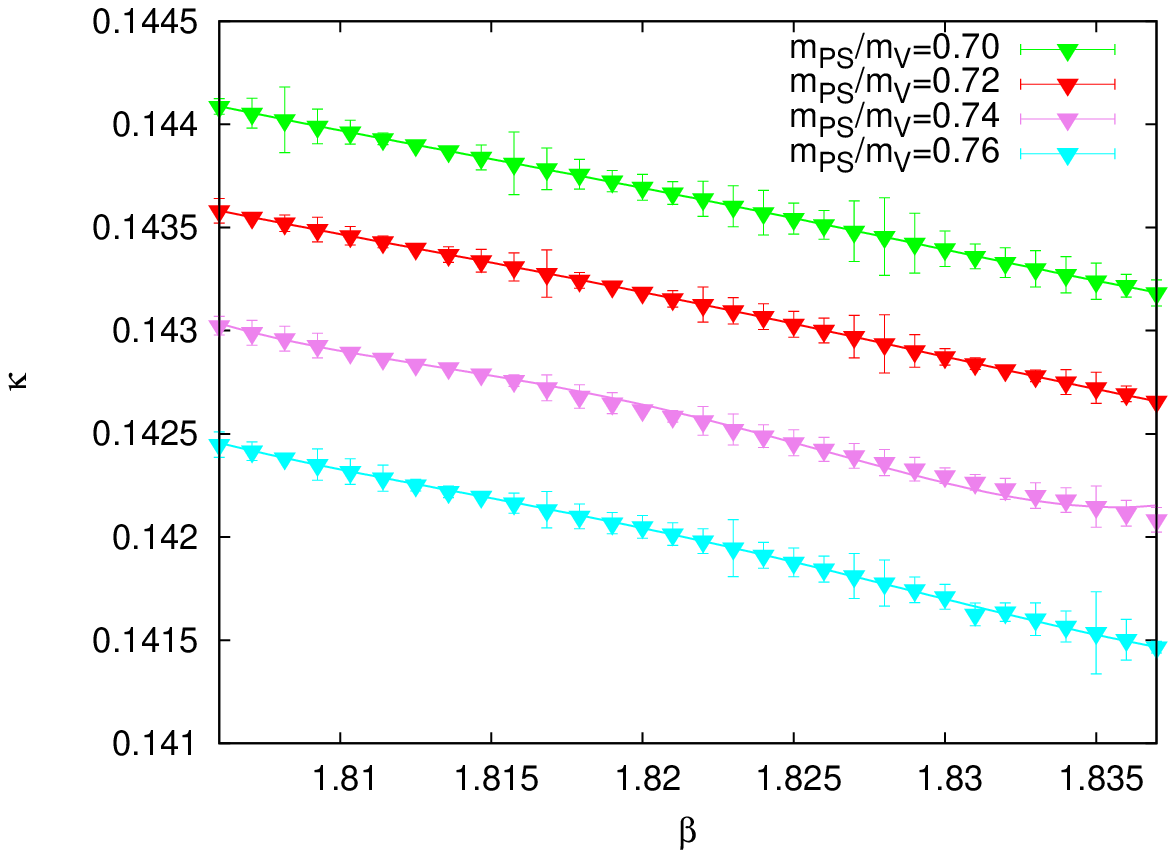}
\caption{The lines of constant physics at $m_{\rm PS}/m_{\rm V}=0.70$, 0.72, 0.74, and 0.76
in the $(\beta, \kappa)$ plane.}
\label{fig:lcp}
\end{minipage}
\end{figure}

\begin{figure}[tb]
\centering
\includegraphics[width=80mm]{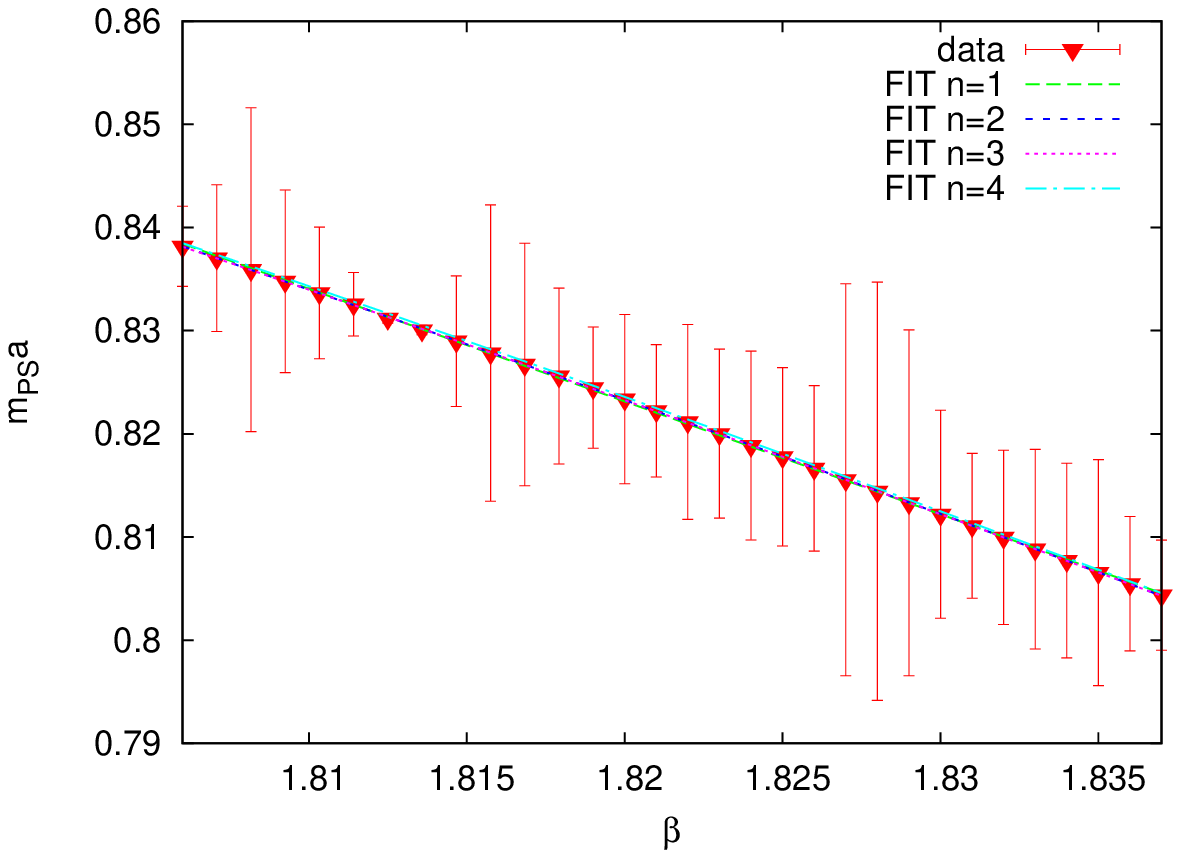}
\includegraphics[width=80mm]{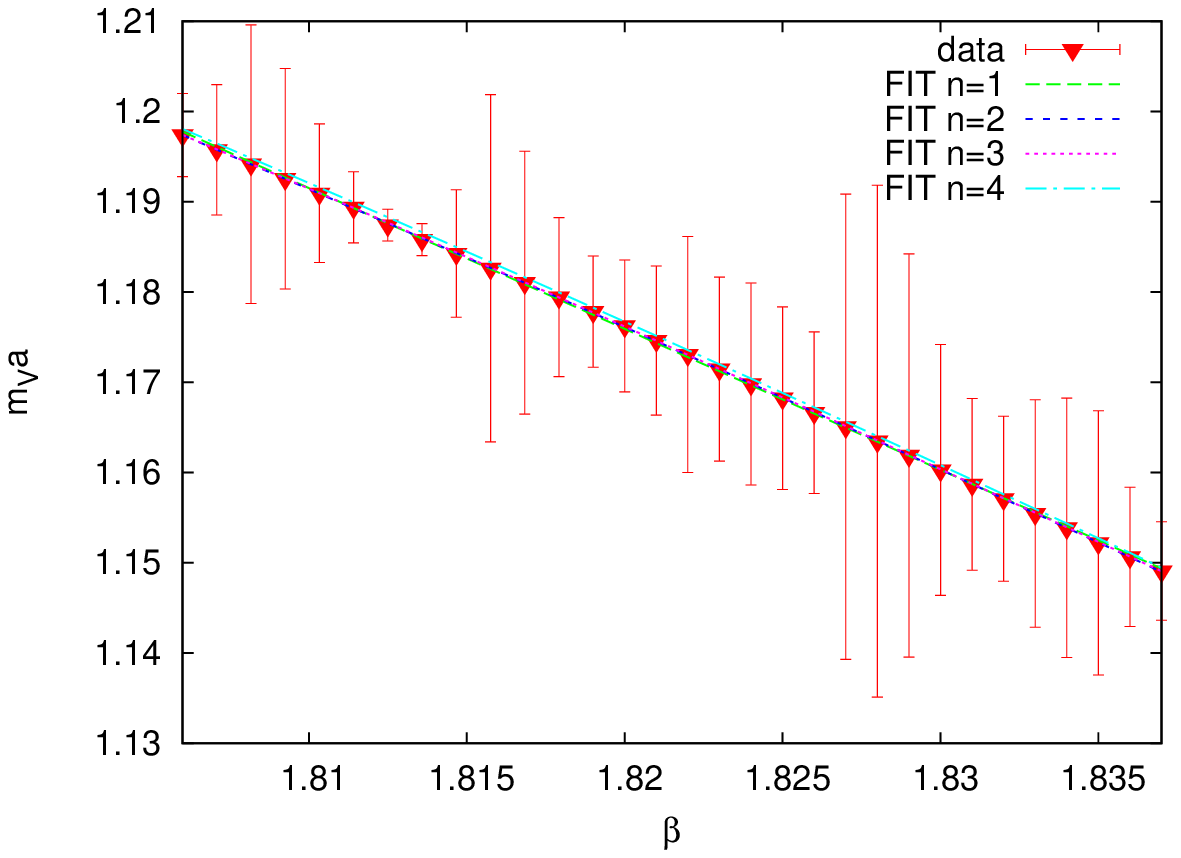}
\caption{The $\beta$-dependence of $m_{\rm PS} a$ (left) and 
$m_{\rm V} a$ (right)  on the line of constant physics at $m_{\rm PS}/m_{\rm V} =0.70$. 
Results of $n$-th order polynomial fits are also shown.
}
\label{fig:avsbeta70}
\end{figure}

\begin{figure}[tb]
\centering
\includegraphics[width=80mm]{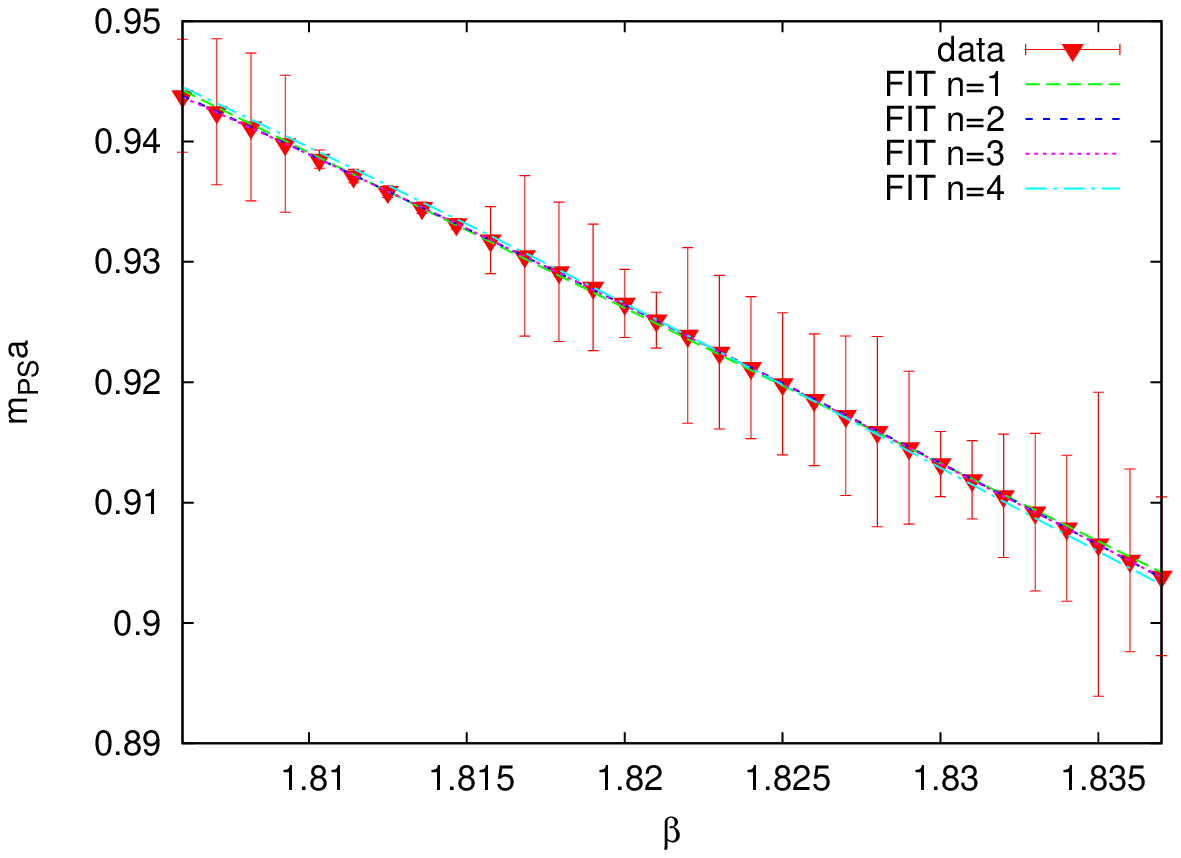}
\includegraphics[width=80mm]{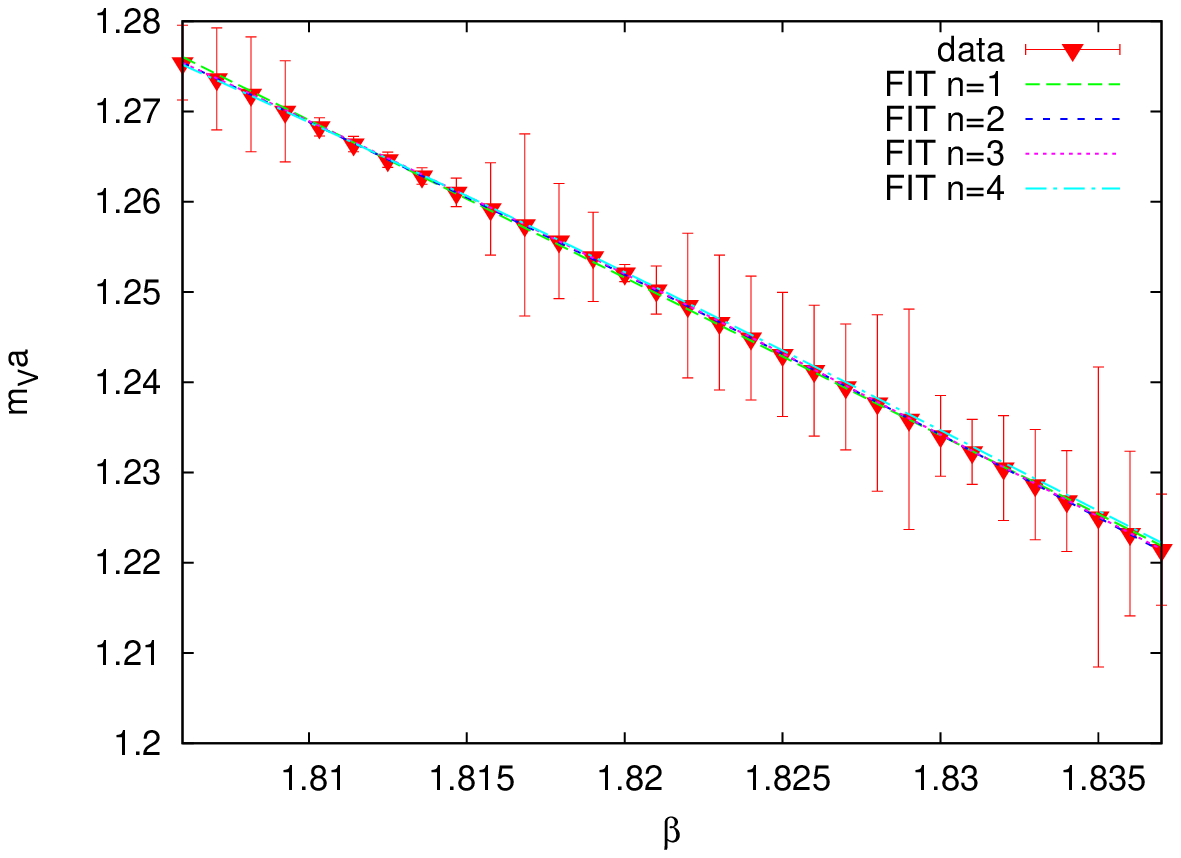}
\caption{The same as Fig.~\ref{fig:avsbeta70} but at $m_{\rm PS}/m_{\rm V} =0.74$. 
}
\label{fig:avsbeta74}
\end{figure}

\begin{figure}[tb]
\centering
\includegraphics[width=80mm]{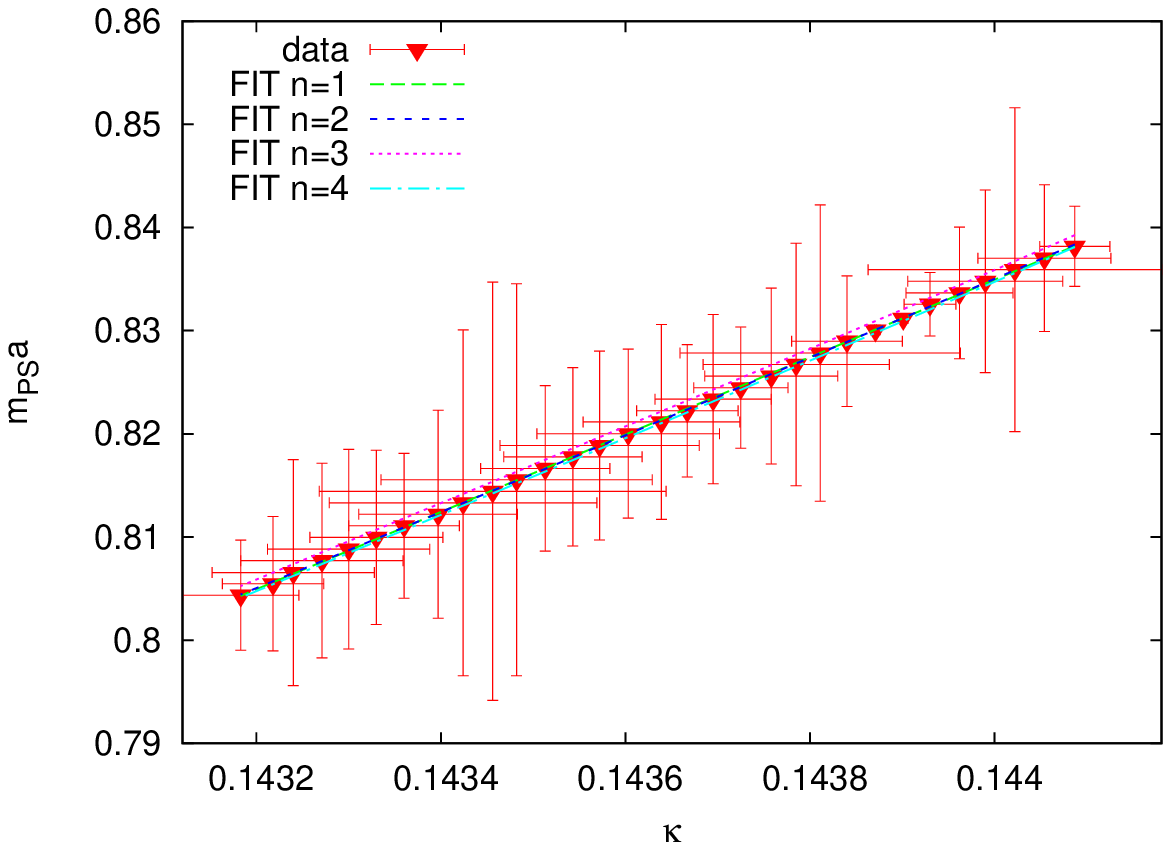}
\includegraphics[width=80mm]{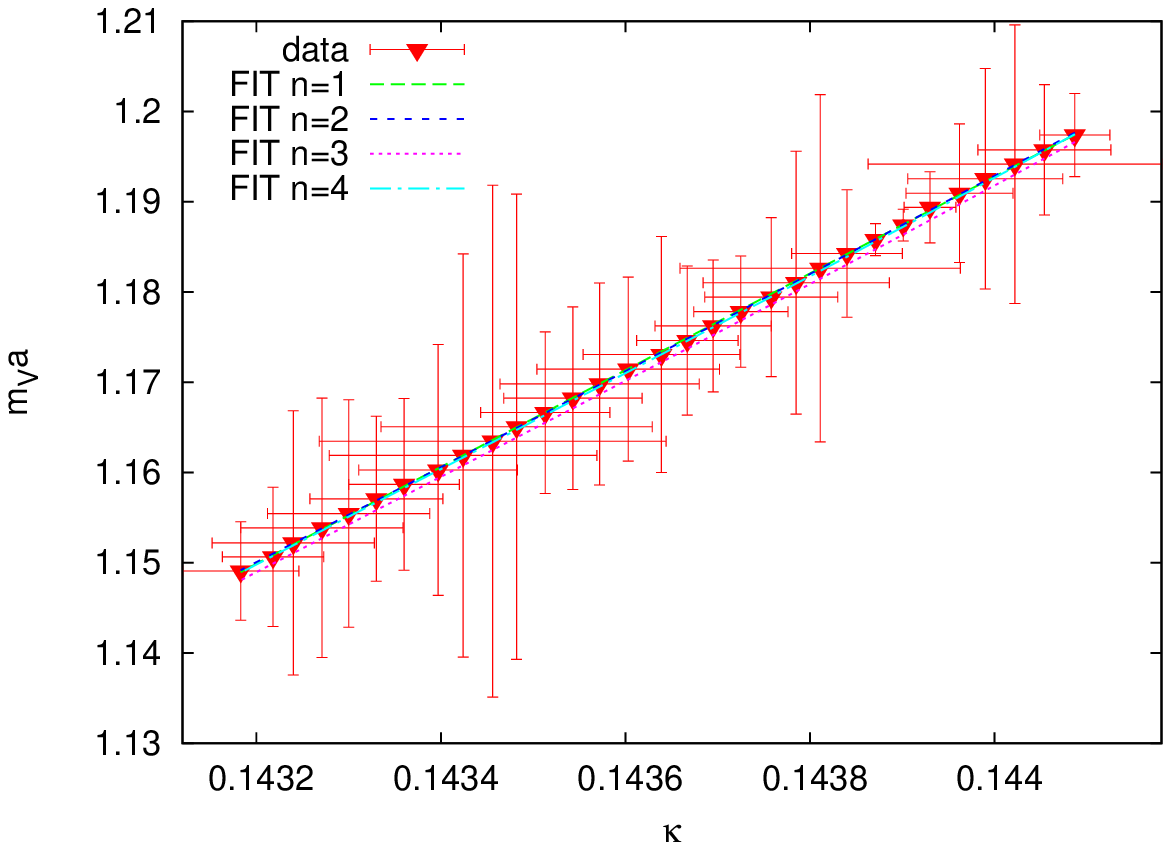}
\caption{The $\kappa$-dependence of $m_{\rm PS} a$ (left) and 
$m_{\rm V} a$ (right) on the line of constant physics at $m_{\rm PS}/m_{\rm V} =0.70$. 
Results of $n$-th order polynomial fits are also shown.
}
\label{fig:avskappa70}
\end{figure}

\begin{figure}[tb]
\centering
\includegraphics[width=80mm]{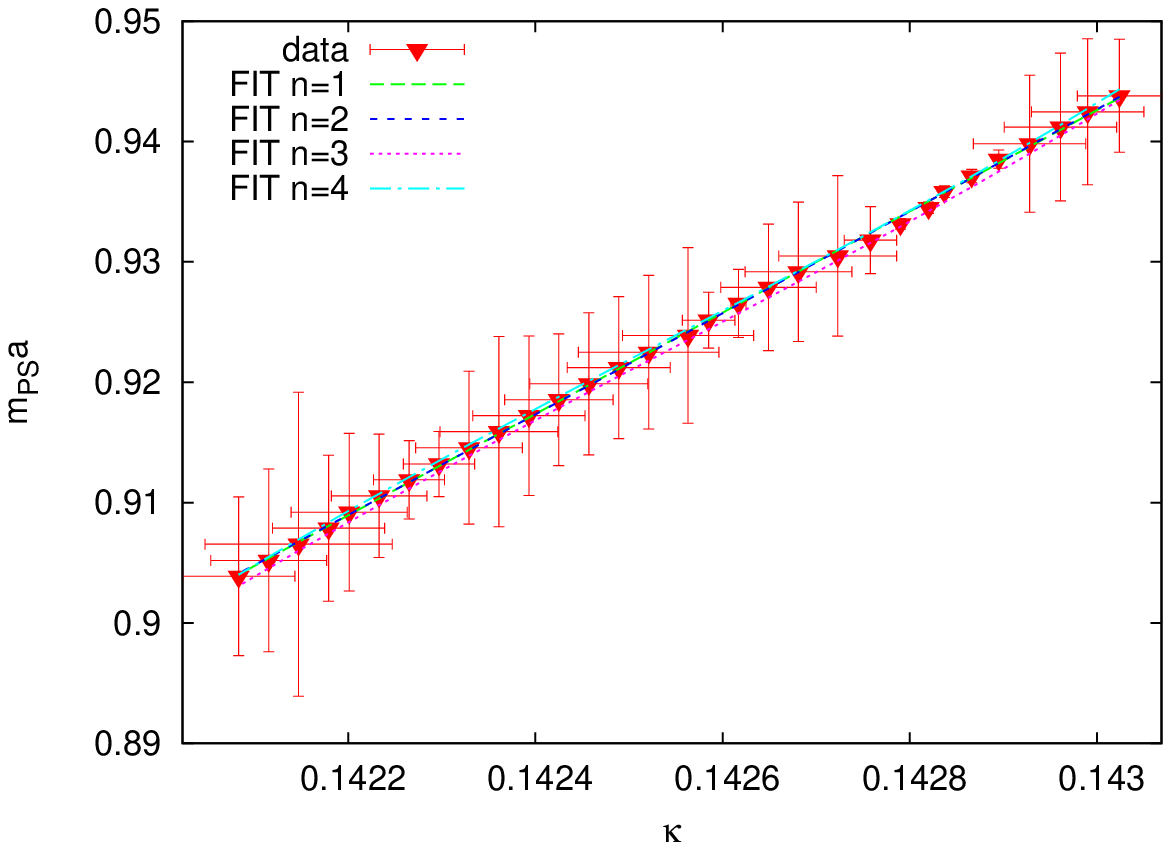}
\includegraphics[width=80mm]{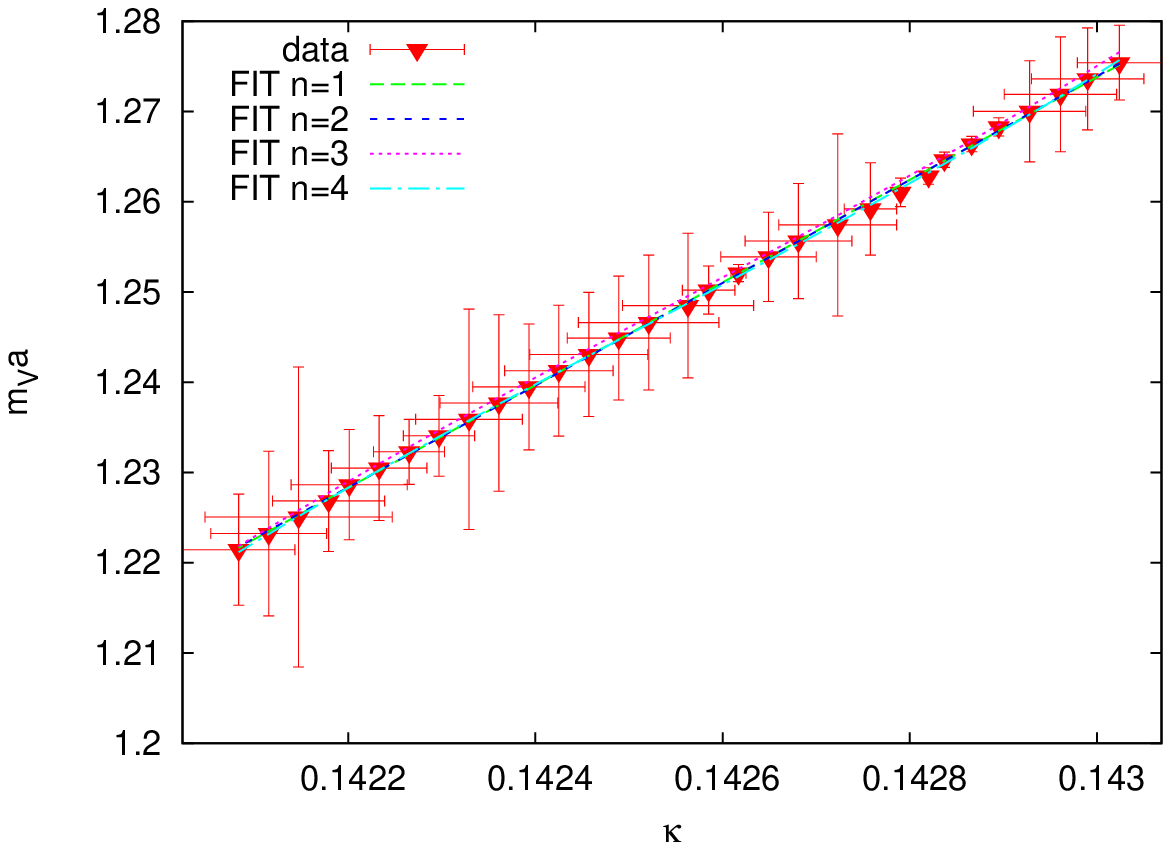}
\caption{The same as Fig.~\ref{fig:avskappa70} but at  $m_{\rm PS}/m_{\rm V} =0.74$. 
}
\label{fig:avskappa74}
\end{figure}

\begin{figure}[tb]
\centering
\includegraphics[width=80mm]{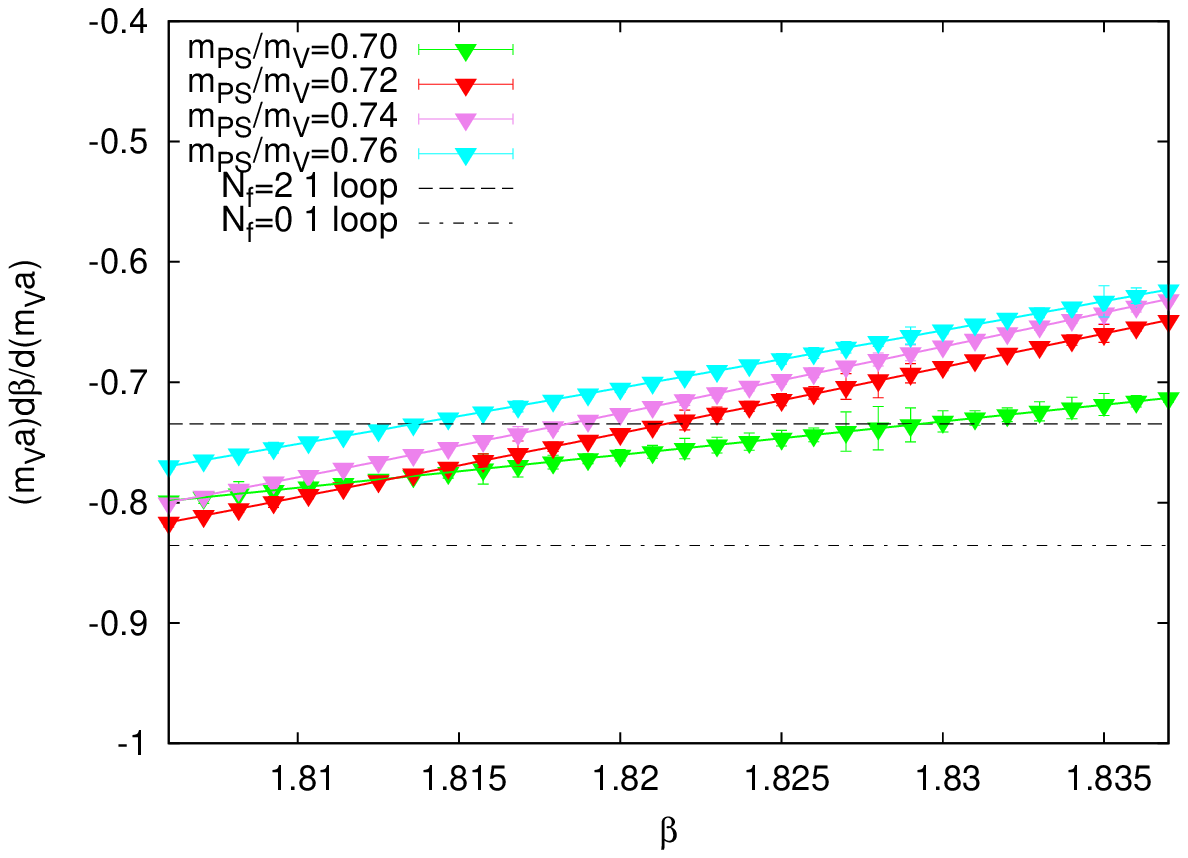}
\includegraphics[width=80mm]{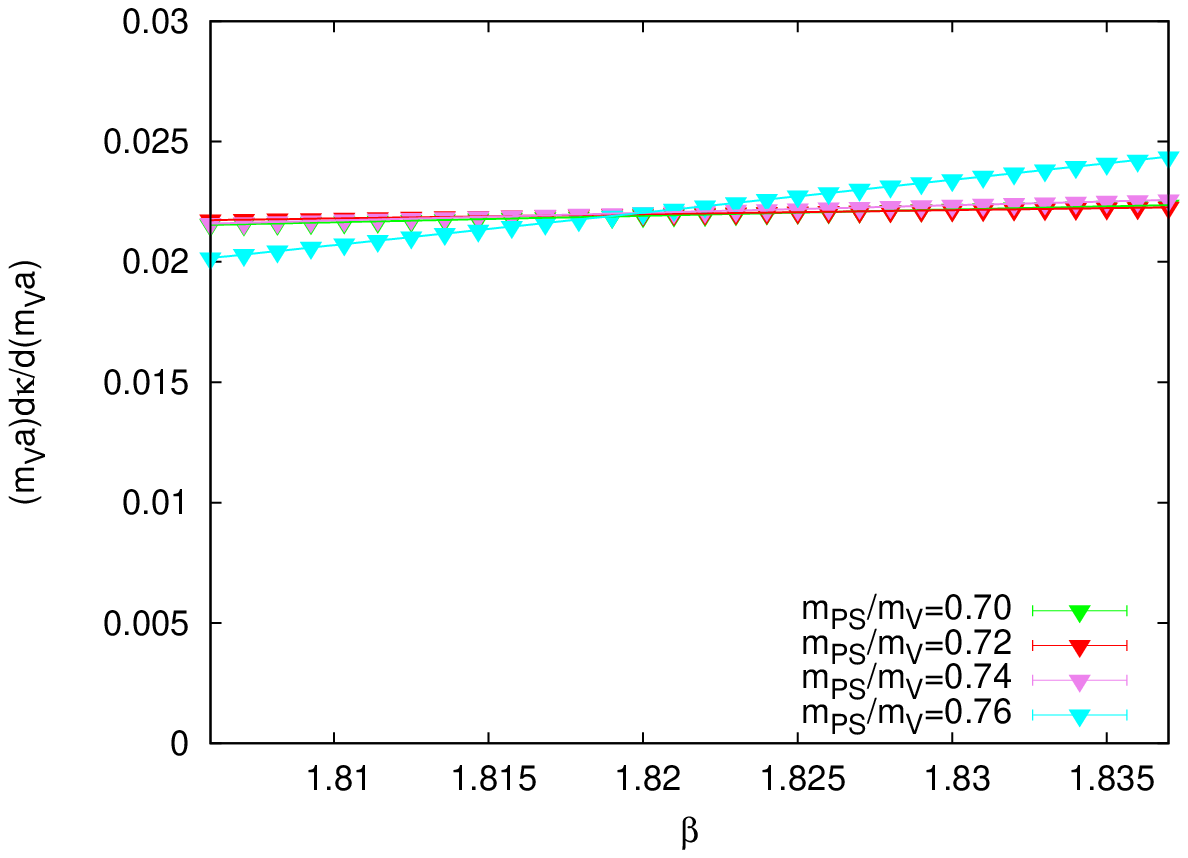}
\caption{Beta functions
$a(d \beta /da)$ (left) and $a(d \kappa /da)$ (right) determined in terms of the quadratic fits ($n=2$) of $m_{\rm V} a$.
The dashed lines are the one-loop perturbative values of $a (d\beta /d a)$ in QCD with zero and two flavors of massless quarks.
}
\label{fig:betafnc}
\end{figure}

\begin{figure}[tb]
\centering
\includegraphics[width=80mm]{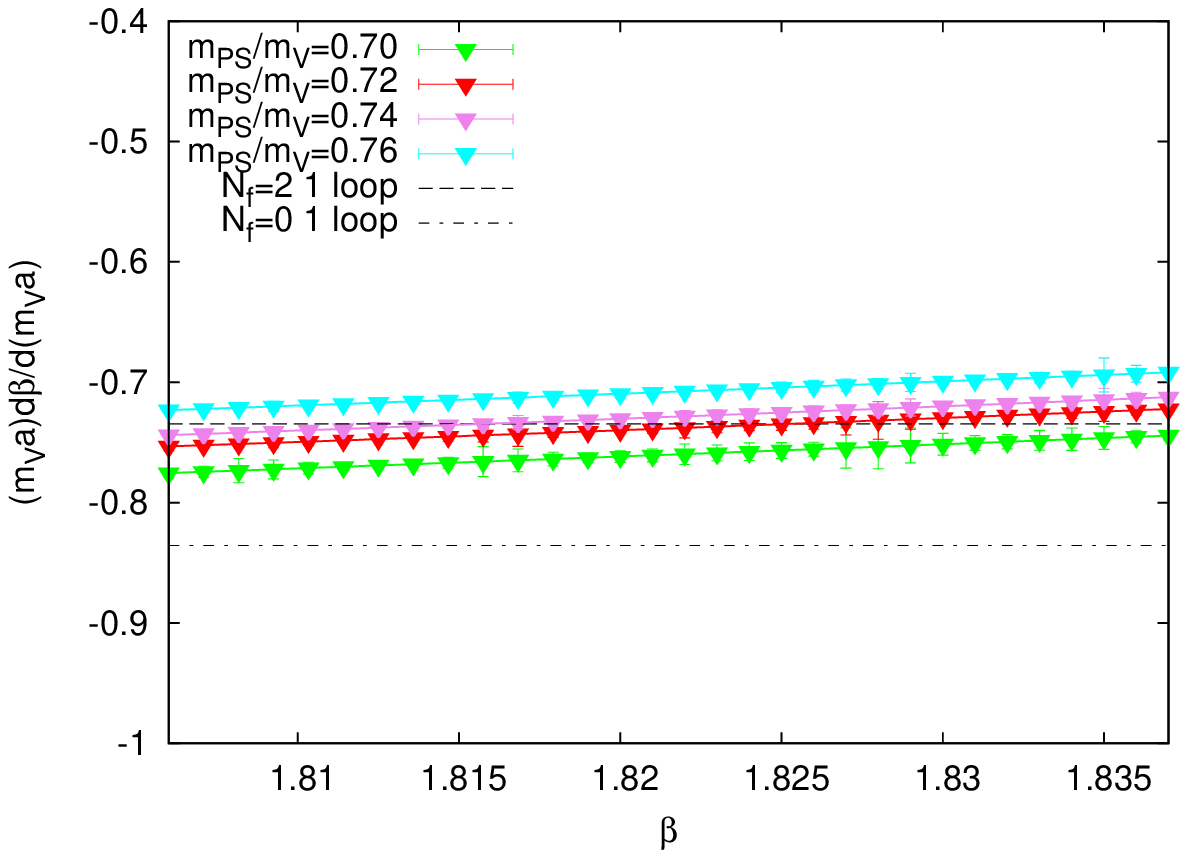}
\includegraphics[width=80mm]{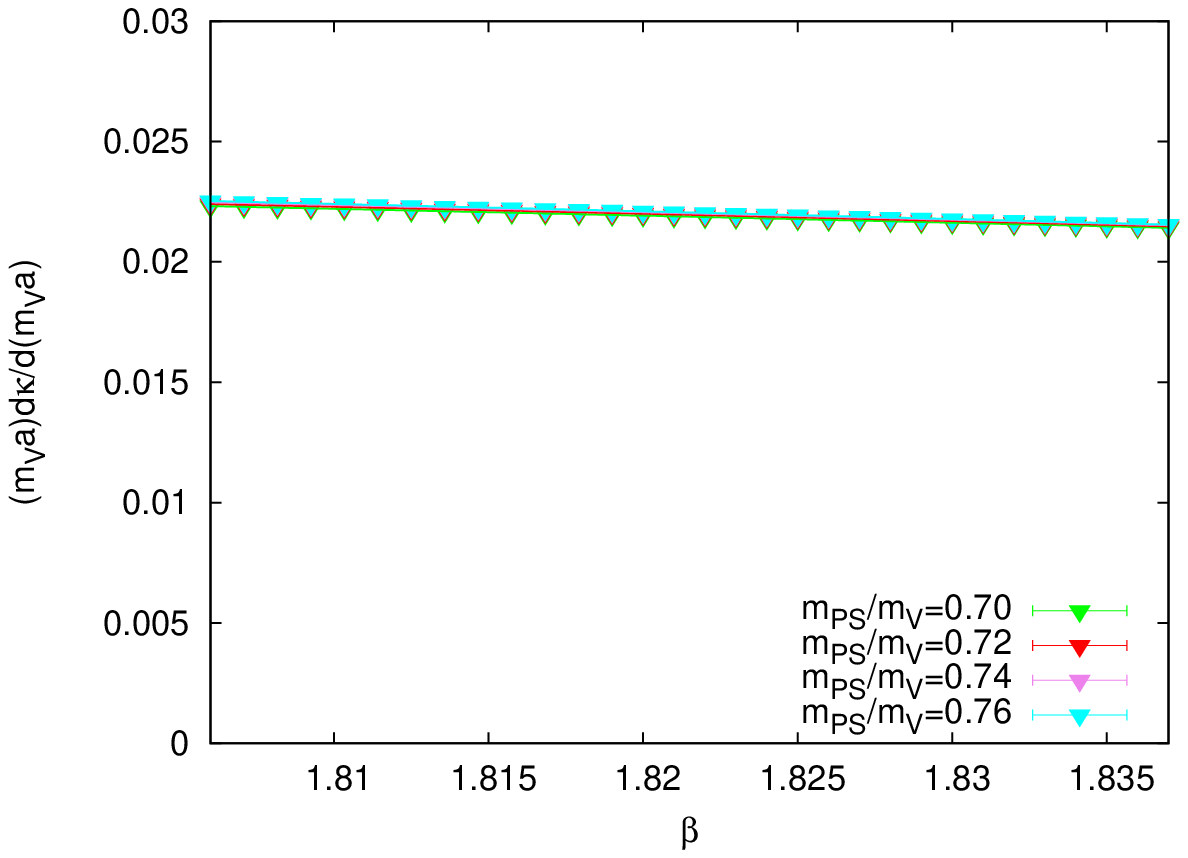}
\caption{The same as Fig.~\ref{fig:betafnc} but by the linear fits ($n=1$).
}
\label{fig:betafnc1}
\end{figure}

Because the multipoint reweighting method enables us to compute observables as continuous functions of $\beta$ and $\kappa$, it is useful to determine the lines of constant physics in the coupling parameter space as well as the beta functions, $a (d \beta / d a)$ and $a (d K / d a)$.
These quantities are needed in a calculation of the equation of state.
To calculate them, we perform simulations at 30 simulation points 
(all the combinations of $\beta= \{ 1.806$, 1.8125, 1.819, 1.825, 1.831, 1.837\} and 
$\kappa= \{ 0.14000$, 0.14125, 0.14250, 0.14300, 0.14400\})
on a $16^4$ lattice. 
We then combine the configurations at the 30 simulation points by the multipoint reweighting method. 
We determine $\ln \det M$ as a function of $\kappa$ using the interpolation method discussed in Sec.~\ref{sec:lndetM}.
For this calculation, we compute the derivatives of $\ln \det M$ by the random noise method with the number of random vectors $N_{\rm noise} =5$ at the five $\kappa$ points. 

In this study, we define the lines of constant physics by fixing the dimensionless ratio of pseudoscalar and vector meson masses, $m_{\rm PS}/m_{\rm V} = m_{\rm PS} a / m_{\rm V} a$, in the $(\beta,\kappa)$ space, where $a$ is the lattice spacing.
Along a line of constant physics thus defined, $a$ varies as we change $\beta$ or $\kappa$. 
The beta functions are defined as the derivatives of $\beta$ and $\kappa$ by $a$ along lines of constant physics. 

To determine the meson masses as functions of $\beta$ and $\kappa$, 
we need meson correlation functions ${\cal G}(t)$ at various $\beta$ and $\kappa$.
Because the computational cost for these correlation functions is relatively low, we compute them at more points of $\beta$ and $\kappa$ than the simulation points using the multipoint reweighting method.
For $\beta$, we choose 31 points, inserting five additional points between each two succeeding simulation points at $\beta= 1.806$, 1.8125, 1.819, 1.825, 1.831, and 1.837.
For $\kappa$, we choose 43 points at $\kappa=0.1400$, 0.1401, $\cdots$, 0.1442. 
(At $\kappa > 0.1442$, we cannot always find a stable plateau in the effective mass plot discussed below.)
At each measurement point, we measure ${\cal G}(t)$ on 200 configurations every 10 trajectories after thermalization.
We also average over 8 different source points. 

We calculate $m_{\rm PS} a$ and $m_{\rm V} a$ by a $\cosh$ fit of the meson correlation function in the range $t/a=5$--8, when a plateau of effective mass is identified in the range. 
The results of $m_{\rm PS} a$ and $m_{\rm V} a$ are shown in the left and right panels of Fig.~\ref{fig:mass}, respectively.
The errors are estimated by the jackknife method.
The mass ratio $m_{\rm PS}/m_{\rm V}$ is shown in Fig.~\ref{fig:pirho}.

At each $\beta$, we interpolate $m_{\rm PS}/m_{\rm V}$ as a function of $\kappa$ to determine the lines of constant physics for $m_{\rm PS}/m_{\rm V}= 0.70$, 0.72, 0.74 and 0.76.
The results are shown in Fig.~\ref{fig:lcp}.
Because $\kappa$ is more sensitive to $m_{\rm PS}/m_{\rm V}$ than $\beta$, the values of $\kappa$ for a line of constant physics greatly vary with $m_{\rm PS}/m_{\rm V}$.

Along a line of constant physics, $m_{\rm PS} a$ and $m_{\rm V} a$ vary with $\beta$ or $\kappa$. 
The $\beta$ and $\kappa$ dependences of these masses at $m_{\rm PS}/m_{\rm V}=0.70$ and 0.74 are plotted in Figs.~\ref{fig:avsbeta70} and \ref{fig:avsbeta74} and  Figs.~\ref{fig:avskappa70} and \ref{fig:avskappa74}, respectively. 
The results at $m_{\rm PS}/m_{\rm V}=0.72$ and 0.76 are similar.
We note that, although the errors estimated at each point vary, the central values form quite smooth curves. 
This may be due to the correlation among different points by the reweighting procedure.

In Figs.~\ref{fig:avsbeta70}-\ref{fig:avskappa74}, results of $n$ th order polynomial fits are shown by dotted lines for $n=1$-4. 
We find that the masses are well fitted with $n=1$ or 2 in the range of $\beta$ and $\kappa$ that we study.

From the fit functions, we can calculate the derivatives, $(m_{\rm PS} a) \,d b /d (m_{\rm PS} a)$ and $(m_{\rm V} a) \,d b /d(m_{\rm V} a)$ with $b=\beta$ and $\kappa$, along the lines of constant physics. 
Because both $m_{\rm PS}$ and $m_{\rm V}$ are constant on a line of constant physics, we expect 
\[
(m_{\rm PS} a) \frac{d b}{d (m_{\rm PS} a)}
= (m_{\rm V} a) \frac{d b}{d (m_{\rm V} a)}
= a \frac{d b}{d a},
\hspace{5mm}
b=\beta,\,\kappa .
\]
We confirm that the beta functions in terms of $m_{\rm PS} a$ and $m_{\rm V} a$ are almost indistinguishable from each other: 
The differences are at most 0.15\% in the range of coupling parameters that we study and are at most 14\% of the statistical errors.
In the following, we adopt $m_{\rm V} a$ for the scale.

Consulting Figs.~\ref{fig:avsbeta70}-\ref{fig:avskappa74}, we adopt the quadratic fits ($n=2$) of $m_{\rm V} a$ for the calculation of the beta functions $a(d \beta /da)$ and $a(d \kappa /da)$. 
The results are shown in Fig.~\ref{fig:betafnc}.
The errors shown are statistical only.
Recall that, because the data at different coupling parameters are correlated due to the reweighting procedure, the statistical errors of the beta functions turn out to be much smaller than those from a naive impression of the meson mass plots.
To get an idea about the magnitude of systematic errors due to the fit ansatz of $m_{\rm V} a$, we also show the results with linear ($n=1$) ansatz in Fig.~\ref{fig:betafnc1}. 

In the left panel of Fig.~\ref{fig:betafnc},  the results of one-loop perturbation theory for $a(d \beta /d a)$ with zero and two flavors of massless quarks are shown by dot-dashed and dashed lines, respectively.
Taking into account the systematic errors, our results are approximately consistent with the two-flavor perturbative value.
On the other hand, we expect that $a(d \kappa /d a)$ approaches zero in the large $\beta$ limit.
Such tendency is not visible yet in the right panel of Fig.~\ref{fig:betafnc}.

\section{Conclusions and outlook}
\label{sec:conclusion}

We studied the multipoint reweighting method in a multidimensional parameter space to avoid the overlap problem.
Performing simulations in two-flavor QCD with Iwasaki's improved gauge action and improved clover quark action, 
we find that the overlap problem can be avoided by appropriately combining configurations at different simulation points by the multipoint reweighting method.
We have further shown that the method is useful in calculating the line of constant physics as well as the beta functions, which are required in the evaluation of thermodynamic properties such as the equation of state.
Extending the multipoint reweighting method to the reweighting study on anisotropic lattices \cite{SIK98}, 
we can also calculate the Karsch coefficients \cite{Karsch82} which are required in the evaluation of the equation of state by the differential method .

Our final objective is to carry out a study of finite density QCD using the multipoint reweighting method. 
In our previous study in the heavy quark region \cite{whot14,whot11}, 
we found that the leading effects of the chemical potential can be absorbed by a shift of coupling parameters 
and the overlap problem is avoided by keeping these shifted parameters constant. 
We expect that a similar shift to absorb the main effects of the chemical potential also exists at lighter quark masses. 
Combining with the multipoint reweighting method, we may be able to investigate the phase structure of finite density QCD at lighter quark masses, maximally avoiding the sign problem.

\section*{Acknowledgments}
This work is in part supported by JSPS KAKENHI Grant
No.\ 26400244, No.\ 26400251, No.\ 26287040, No.\ 15K05041, and by the 
Large Scale Simulation Program of High Energy Accelerator
Research Organization (KEK) No.\ 13/14-21 and No.\ 14/15-23.


\end{document}